\documentclass[%
reprint,
twocolumn,%
 amsmath,amssymb,
aps,
prx,
]{revtex4-2}
\usepackage{xcolor}
\usepackage{graphicx}%
\usepackage{dcolumn}%
\usepackage{bm}%
\usepackage{physics}
\usepackage{bbm}
\usepackage{siunitx}
\usepackage{amsmath}
\usepackage{dsfont}
\usepackage{array,booktabs}
\newcolumntype{C}{>{$\displaystyle}c<{$}}
\usepackage{chngcntr}
\usepackage{float}
\usepackage{libertine}
\definecolor{NewBlue}{rgb}{0, 0, 0.41}
\definecolor{NewRed}{rgb}{0.6, 0.07, 0.07}
\usepackage[colorlinks,
    linkcolor=NewBlue,
    citecolor=NewBlue,
    urlcolor=NewBlue]{hyperref}

\newcommand{\wR}{\Omega}
\newcommand{\sens}{v_{\mathrm{min}}}
\newcommand{\effrabi}{\tilde{\Omega}}
\newcommand{\effrabimax}{\tilde{\Omega}_{\mathrm{max}}}
\newcommand{\ide}{\mathbb{I}}
\newcommand{\wrf}{\omega_\mathrm{RF}}

\newcommand{\df}{\delta_\phi}
\newcommand{\dzero}{\Delta_0}

\newcommand{\wzero}{\omega_0}

\newcommand{\wf}{\omega_\phi}
\newcommand{\wmean}{\bar{\omega}}
\newcommand{\Df}{\Delta_\phi}
\newcommand{\ws}{w_\mathrm{s}}
\newcommand{\Uideal}{U_{\mathrm{i}}}
\newcommand{\Ucalculated}{U_{\mathrm{c}}}

\renewcommand{\appendixname}{APPENDIX}

\begin{document}

\title{Improved Electron-Nuclear Quantum Gates for Spin Sensing and Control}

\author{H. B. van Ommen$^{1,2}$}
\thanks{These authors contributed equally}
\author{G. L. van de Stolpe$^{1,2}$}
\thanks{These authors contributed equally}
\author{N. Demetriou$^{1,2}$}
\author{H. K. C. Beukers$^{1,2}$}
\author{J. Yun$^{1,2}$}
\author{T. R. J. Fortuin$^{1,2}$}
\author{M. Iuliano$^{1,2}$}
\author{A. R.-P. Montblanch$^{1,2}$}
\author{R. Hanson$^{1,2}$}
\author{T. H. Taminiau$^{1,2}$}
\email{t.h.taminiau@tudelft.nl}

\affiliation{$^1$QuTech, Delft University of Technology, PO Box 5046, 2600 GA Delft, The Netherlands}%

\affiliation{$^2$Kavli Institute of Nanoscience Delft, Delft University of Technology,
PO Box 5046, 2600 GA Delft, The Netherlands}

\date{\today}%

\begin{abstract}
The ability to sense and control nuclear spins near solid-state defects might enable a range of quantum technologies. Dynamically Decoupled Radio-Frequency (DDRF) control offers a high degree of design flexibility and long electron-spin coherence times. However, previous studies considered simplified models and little is known about optimal gate design and fundamental limits. Here, we develop a generalised DDRF framework that has important implications for spin sensing and control. Our analytical model, which we corroborate by experiments on a single NV center in diamond, reveals the mechanisms that govern the selectivity of gates and their effective Rabi frequencies, and enables flexible detuned gate designs. We apply these insights to numerically show a 60x sensitivity enhancement for detecting weakly coupled spins and study the optimisation of quantum gates in multi-qubit registers. These results advance the understanding for a broad class of gates and provide a toolbox for application-specific design, enabling improved quantum control and sensing. 
\end{abstract}

\maketitle

\newpage
\section{Introduction}
Sensing and controlling nuclear spins in the vicinity of optically active solid-state defects, such as the nitrogen vacancy (NV) center in diamond, has opened up various opportunities in the fields of quantum sensing and quantum information processing \cite{lovchinsky_nuclear_2016,pompili_realization_2021,abobeih_atomicscale_2019, randall_manybody_2021}. Sensing nuclear spins outside the host crystal might bring chemical structure determination to the single-molecule level \cite{lovchinsky_nuclear_2016, abobeih_atomicscale_2019, cujia_parallel_2022, budakian_roadmap_2024}. More strongly coupled nuclear spins inside the host material can be used for quantum information processing, for which advances in the number of available qubits \cite{vandestolpe_mapping_2024}, in gate fidelities \cite{xie_99_2023, bartling_universal_2024} and in the possibility to connect systems via an optical interface \cite{pompili_realization_2021, humphreys_deterministic_2018} have led to proof-of-principle demonstrations of increasing complexity \cite{hermans_qubit_2022, abobeih_atomicscale_2019, randall_manybody_2021}.

Central to these developments has been the ability to sense and control nuclear spins using the defect's electron spin through the hyperfine interaction \cite{taminiau_detection_2012, vandersar_decoherenceprotected_2012, bradley_tenqubit_2019, abobeih_onesecond_2018, bartling_entanglement_2022,schwartz_robust_2018}. In particular, dynamical decoupling (DD) protocols have been used to detect nuclear magnetic resonance signals \cite{staudacher_nuclear_2013, cujia_tracking_2019} and allow for selective, universal nuclear spin control \cite{taminiau_universal_2014}. Compared to traditional DD sensing, the recently developed DDRF sequence \cite{bradley_tenqubit_2019}, which combines DD with radio-frequency (RF) pulses, unlocks additional sensing and control directions (Fig. \ref{fig:DDRF_intro}a) and offers increased flexibility for gate optimisation \cite{bartling_universal_2024}. These advantages have helped enable the sensing of large nuclear spin clusters \cite{abobeih_atomicscale_2019, vandestolpe_mapping_2024}, extend the number of nuclear spins available to defect centers for information processing \cite{bradley_tenqubit_2019, abobeih_faulttolerant_2022, parthasarathy_scalable_2023}, and realise high two-qubit gate fidelities $(>99.9\%)$ \cite{bartling_universal_2024}.

In this work, we introduce a generalised version of the DDRF framework, enhancing the sequence's performance for nuclear sensing and control, as well as revealing important limitations on sensitivity and selectivity. We derive analytical expressions that give a more complete description of the electron-nuclear dynamics compared to previous work \cite{bradley_tenqubit_2019} and verify their predictions experimentally using a single NV center and its surrounding $^{13}$C nuclear spins. Based on these insights, we modify the DDRF sequence to optimise the effective electron-nuclear interaction strength and mitigate crosstalk of quantum gates. These results have applications in the field of nano-NMR \cite{budakian_roadmap_2024,lovchinsky_nuclear_2016,abobeih_atomicscale_2019} and provide a comprehensive toolbox for designing quantum gates in multi-qubit electron-nuclear spin systems \cite{abobeih_faulttolerant_2022,bradley_tenqubit_2019}. 

\section{Decoherence-protected radio-frequency quantum gates} \label{sec:DDRF_intro}
\begin{figure*}
  \includegraphics[width=1 \textwidth]{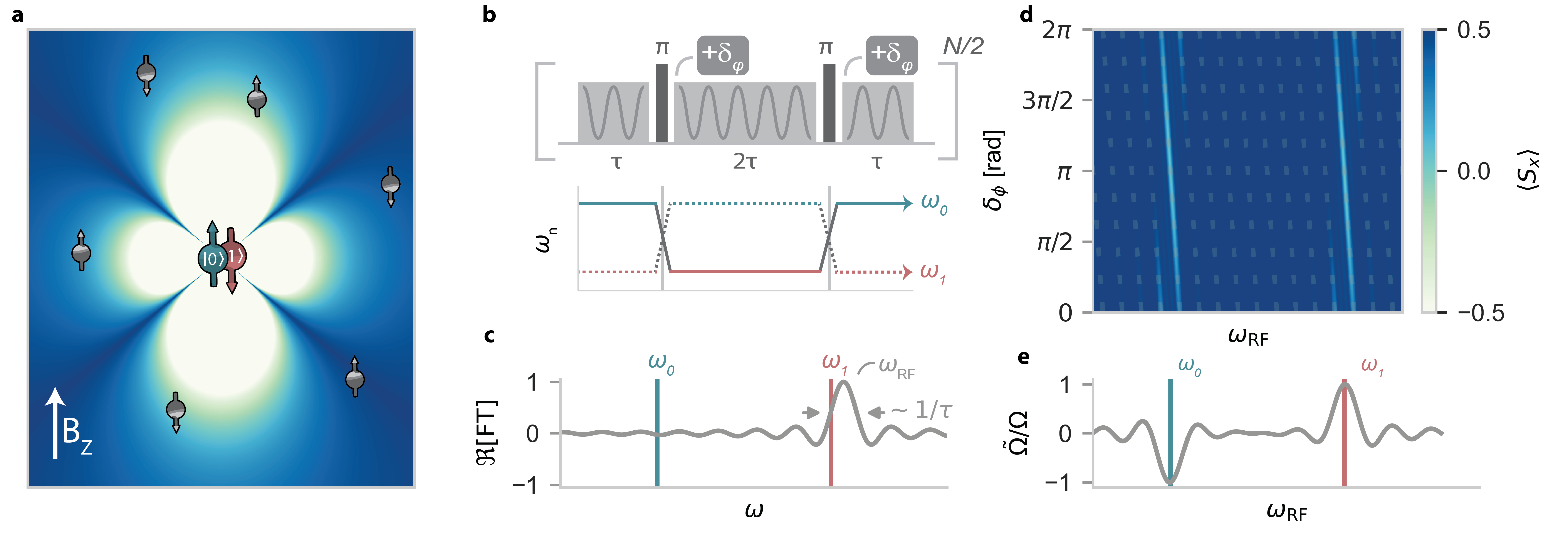}
  \caption{\emph{The DDRF framework.} a) Schematic of the system considered here, comprising a single electron spin (red/blue colors denote the spin state), surrounded by a non-uniform distribution of nuclear spins (grey). The background brightness corresponds to the hyperfine ($A_\parallel$) frequency shift $\Delta$ induced by the electron spin. $B_z$ is the external magnetic field. b) DDRF sequence, where RF driving (with phase updates $\df$) of the nuclear spin (grey) is interleaved with decoupling pulses on the electron spin (black). The bottom row indicates the nuclear spin's precession frequency ($\omega_\mathrm{n}$), for initial electron-spin state $\ket{0}$ (solid line) or $\ket{1}$ state (dotted line). The total sequence time is $T=2N\tau$. c) Schematic showing the Fourier transform (FT) of a single RF pulse applied at frequency $\wrf = \omega_1 + \Delta_1$. The bandwidth of the pulse is inversely proportional to the pulse length ($\propto 1/(2\tau)$). d) Simulated nuclear-spin spectroscopy, where the electron spin (starting in $\tfrac{\ket{0} + \ket{1}}{\sqrt{2}}$) is used to sense an unpolarised nuclear spin by applying the DDRF sequence of (b). Brighter colors indicate a drop in electron coherence ($\expectationvalue{S_x}$), observed when $\df$ matches Eq. \ref{eq:phase_update_rule} (grey dashed line). e) Effective nuclear-spin Rabi frequency $\effrabi$ (Eq. \ref{eq:eff_rabi}) of the DDRF sequence, as a fraction of the bare Rabi frequency $\Omega$, when following the phase increment $\df$ (Eq. \ref{eq:phase_update_rule}, dashed line in d). A significant electron-nuclear interaction is possible over a range of RF frequencies ($\wrf \neq \omega_1$), albeit with a lower effective interaction strength $\effrabi$.
  \label{fig:DDRF_intro}
  }
\end{figure*}

We first describe the DDRF gates. Compared to the original description \cite{bradley_tenqubit_2019}, we present a generalized framework, explicitly including the off-resonant driving of nuclear-spin transitions. We will show that this refinement has important implications for the performance of quantum gates and sensing sequences implemented with DDRF control.

We consider an electronic spin interacting with a number of nuclear spins via a hyperfine interaction (see Fig. \ref{fig:DDRF_intro}a). To retain generality for defects with different spin numbers \cite{babin_fabrication_2022,uysal_coherent_2023,higginbottom_optical_2022, debroux_quantum_2021, sipahigil_integrated_2016}, we assume two electron spin states are selected to use as a qubit and describe these as a pseudo spin-$\tfrac{1}{2}$ system spanned by $\ket{0}$ and $\ket{1}$. 

The main challenge for electron-nuclear gates in such systems is that the electron interacts with all nuclear spins, as well as other noise sources, leading to decoherence and crosstalk \cite{bradley_tenqubit_2019, vandersar_decoherenceprotected_2012, hannes_fidelity_2024}. Hence, a well-designed electron-nuclear two-qubit gate or sensing sequence aims to realize a conditional interaction with a selected (group of) target spin(s), while protecting electron coherence by decoupling all other interactions and noise sources.

The DDRF gate consists of a sequence of dynamical decoupling (DD) pulses on the electron spin, interleaved with RF pulses that drive the nuclear spin transitions, as illustrated in Fig. \ref{fig:DDRF_intro}b. We consider sequences of the form $(\tau - \pi  - 2\tau - \pi -\tau)^{N/2}$, with $N$ the number of $\pi$-pulses and $2\tau$ the interpulse delay. The DD sequence aims to decouple the electron spin from the surrounding spins and magnetic field fluctuations, extending the electron-spin coherence \cite{abobeih_onesecond_2018}. The interleaved RF pulses aim to manipulate selected nuclear spins and to re-couple them to the electron spin \cite{bradley_tenqubit_2019}.

In the frame of the electron energy splitting, the Hamiltonian for the electron spin and a single nuclear spin is \cite{bradley_tenqubit_2019} (Appendix \ref{sec:sup_nv_hamiltonian}):

\begin{equation} \label{eq:H_stat}
    H = \bar{\omega} I_z + \Delta \sigma_z I_z \,,
\end{equation}
where $\bar{\omega} = (\omega_0 + \omega_1)/2$ is the mean nuclear-spin frequency, with $\omega_0$ and $\omega_1$ the nuclear spin precession frequencies for electron-spin states $\ket{0}$ or $\ket{1}$, respectively. $\Delta = \omega_0 - \omega_1$ is set by the strength of the electron-nuclear hyperfine interaction \cite{taminiau_detection_2012} (Appendix \ref{sec:sup_nv_hamiltonian}). $\sigma_z$ and $I_z$ are the electronic and nuclear spin-$\tfrac{1}{2}$ operators, respectively. Note that we neglect the anisotropy of the hyperfine interaction (terms such as  $A_\perp\sigma_z I_x$). While this interaction can create complex dynamics and can be used for qubit control \cite{taminiau_universal_2014, taminiau_detection_2012}, the effects can be minimized by applying strong magnetic fields ($\bar{\omega} \gg A_\perp$) and setting the interpulse delay $\tau$ to a multiple of the nuclear spin Larmor period $\tau_L = 2\pi/\omega_L$\cite{bradley_tenqubit_2019}.

The RF pulses selectively drive nuclear spins, recoupling them to the electron-spin. In the interaction picture, the Hamiltonian during the RF pulses for a single nuclear spin is (in the rotating frame at the RF frequency $\wrf$):
\begin{multline}\label{H_rot}
    H_{\mathrm{RF}} = \ket{0}\bra{0} \otimes \Delta_0 I_z 
    + \ket{1}\bra{1} \otimes \Delta_1 I_z \\
    + \ide \otimes \wR (\cos \phi I_x + \sin \phi I_y),        
\end{multline}
with $\Delta_{0} = \wrf - \omega_0$ and $\Delta_{1} = \wrf - \omega_1$  the detunings between the nuclear-spin transition frequencies and the RF frequency, $\phi$ the phase and $\wR$ the Rabi frequency of the (bare) RF drive. 

Because the frequencies $\Delta_0$ and $\Delta_1$ differ by $\Delta$, an RF pulse will generally cause a different nuclear-spin evolution for the $\ket{0}$ and $\ket{1}$ electron states, enabling the construction of conditional two-qubit gates. Similarly, other spins with $\Delta_0', \Delta_1' \neq \Delta_0, \Delta_1$ undergo a different evolution, introducing an element of selectivity between different nuclear spins. Previous work \cite{bradley_tenqubit_2019} assumed that resonant RF driving ($\Delta_1=0$) combined with $\Delta_0 \gg \Omega$ resulted in negligible driving during the electron $\ket{0}$ state, thus neglecting that part of the driving term in $H_{\mathrm{RF}}$. Below, we show that this term cannot generally be neglected due to the broad effective bandwidth of the short RF pulses (small $\tau$) in the DDRF sequence. 

To ensure that the DDRF sequence generates the desired gate, the phase of each RF pulse must be set so that the pulses result in a constructive build-up of rotations on the nuclear spin. This equates to following the phase evolution of the nuclear spin in the frame of the RF frequency. In the decoupling sequence, this is achieved by incrementing the phase of the next RF pulse by a phase-angle $\df$ every time a decoupling $\pi$-pulse is applied on the electron spin (Fig. \ref{fig:DDRF_intro}b). 

In one DDRF block ($N=2$), the nuclear spin accumulates a total phase of $2 \Delta_0 \tau + 2 \Delta_1 \tau$ (up to a correction for the AC-Stark shift, see Appendix \ref{sec:sup_optimal_phase}). By adding a $\pi$ phase shift with each decoupling pulse, the direction of the RF drive is inverted synchronous to the flipping of the electron spin state, creating a conditional electron-nuclear interaction. This gives rise to a resonance condition, satisfied by setting a single-pulse phase increment:
 \begin{equation} \label{eq:phase_update_rule}
    \df = -\left( \Delta_0 + \Delta_1 \right)\tau  \, +\pi ,
\end{equation}
up to multiples of $2\pi$. The dependence of the mean precession frequency during the gate, ($\Delta_0 + \Delta_1)/2$, on the electron-nuclear hyperfine interaction means that this resonance condition provides an additional mechanism for selectivity between different nuclear spins. Importantly, Eq. \ref{eq:phase_update_rule} constitutes a generalisation of the phase-increment resonance considered in previous work (restricted to $\Delta_1 = 0, \Delta_0 = \Delta$) \cite{bradley_tenqubit_2019}.

To quantify the strength of the conditional interaction, we evaluate the unitary of the total DDRF sequence under the Hamiltonian in Eq. \ref{H_rot}, setting the phase increment to Eq. \ref{eq:phase_update_rule}. We assume that the rotation due to an individual RF pulse is small ($\Omega \tau \ll 1$), which is typical for DDRF gates, because the gate's total rotation angle is broken up into $N$ pieces. In this limit, the evolution can be described by a conditional rotation $V_\mathrm{CROT}$ of the nuclear spin (\ref{sec:sup_eff_rabi_derivation}):
\begin{equation}
\label{eq:vcrot}
    V_\mathrm{CROT} = \ket{0}\bra{0} \otimes R_x(N \effrabi \tau )
    + \ket{1}\bra{1} \otimes R_x(-N \effrabi \tau) \, ,
\end{equation}
with an \emph{effective} Rabi frequency given by:
\begin{equation} \label{eq:eff_rabi}
    \effrabi=\wR \left(  \mathrm{sinc}(\Delta_1 \, \tau) - \mathrm{sinc}(\Delta_0 \, \tau) \right) \,,
\end{equation}
where $R_x(\theta) = e^{-i \theta I_x}$, with the $x$ axis set by the phase of the first RF pulse and the sinc function is defined as: $\mathrm{sinc}(x) = 
\sin(x)/x$. Note that previous work neglected off-resonant driving, setting $\Delta_0 \tau \gg 1$ and $\Delta_1 =0 $, so that Eq. \ref{eq:eff_rabi} reduces to $\effrabi = \Omega$ \cite{bradley_tenqubit_2019}.

Setting $\effrabi N\tau  = \pi / 2$ in Eq. \ref{eq:vcrot} results in a fully entangling gate, equivalent to a CNOT up to single qubit rotations. Furthermore, Eq. \ref{eq:eff_rabi} shows that such a gate can be constructed in the neighborhood of the $\omega_0$ and $\omega_1$ frequencies, over a bandwidth given by $\sim 1/\tau$ (see Figs. \ref{fig:DDRF_intro} c-e). Note that this can be understood as the result of evaluating the Fourier transform of an individual RF pulse (applied at $\omega_{\mathrm{rf}}$) at the nuclear-spin transitions $\omega_0$ and $\omega_1$, and that the bandwidth is much larger than would be expected from power broadening due to $\Omega$. 

In the next sections, we first experimentally verify Eqs. \ref{eq:phase_update_rule} and \ref{eq:eff_rabi} by performing DDRF spectroscopy on a single NV center in diamond (Section \ref{sec:generalised_DDRF}). Then, we investigate the weak-coupling regime ($\Delta \tau \lesssim \pi$), for which Eq. \ref{eq:eff_rabi} poses an inherent trade-off between effective interaction strength and electron decoherence (Section \ref{sec:global_window}). Finally, we apply these findings to two applications: sensing a single nuclear spin (Section \ref{sec:sensing_optimisation}) and qubit control in a realistic nuclear-spin quantum register (Sections \ref{sec:gate_optimisation} and \ref{sec:multi_spin_register}).

\section{Generalised DDRF spectroscopy} \label{sec:generalised_DDRF}
\begin{figure}
  \includegraphics[width=1 \columnwidth]{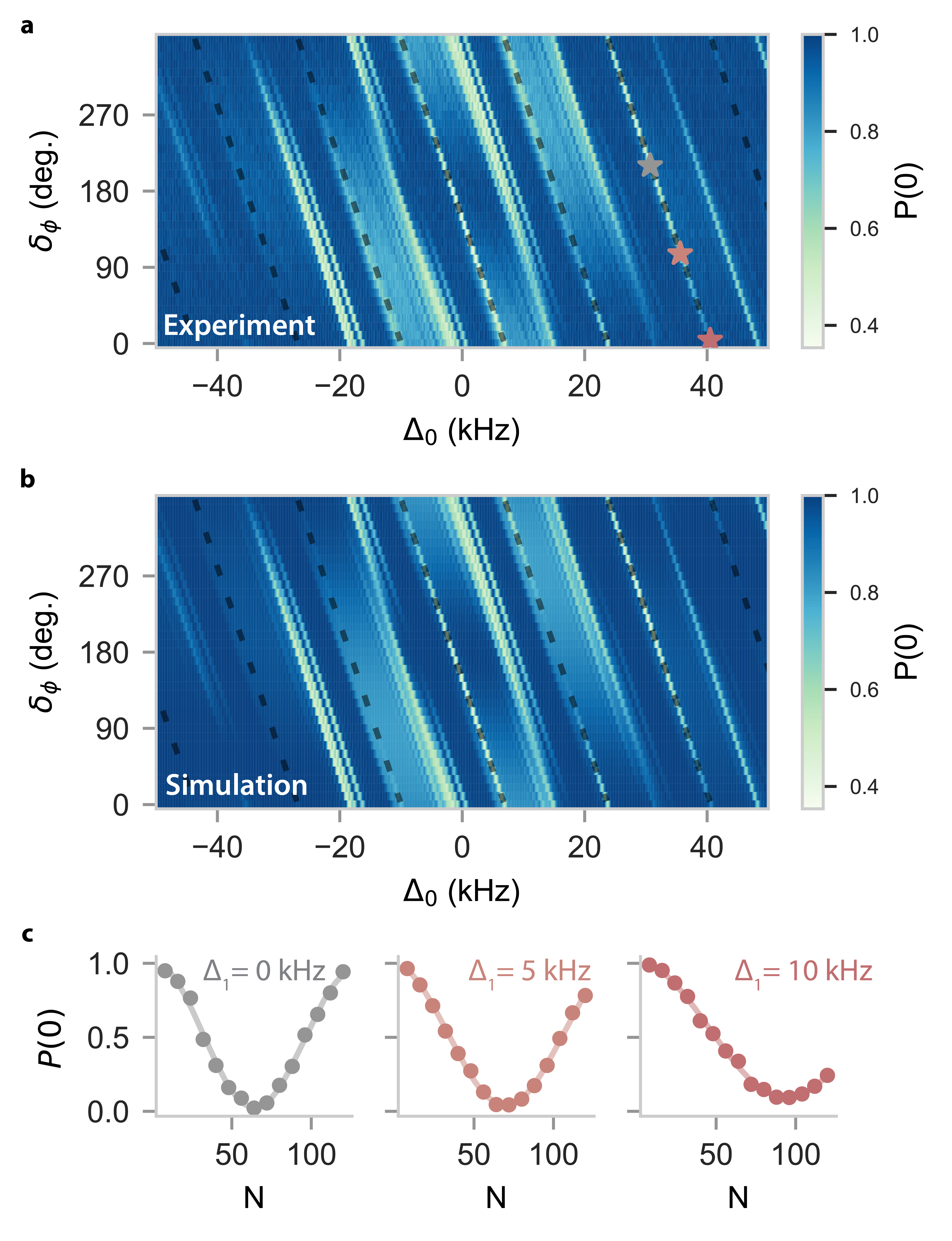}
  \caption{\emph{Generalised DDRF spectroscopy.} a) Experimental data showing DDRF spectroscopy on the nuclear spin environment of an NV center, sweeping the RF frequency $\dzero = \wrf - \wzero$ and the phase update $\df$ ($\tau =$ 29.632us, $N=24$, $\Omega=356$ Hz). Single nuclear spins show up as descending diagonal lines following their frequency-phase resonance (Eq. \ref{eq:phase_update_rule}), while the bath of weakly-coupled spins shows up as a band. The black dashed line indicates the resonance condition of a single spin ($C_0$, $\Delta = \SI{-30.7}{\kilo \hertz}$). Colored stars correspond to the parameters used for the data in (c). $P(0)$ corresponds to the remaining electron spin coherence (Appendix \ref{sec:experimental_methods}). b) Numerical calculation of the DDRF spectroscopy signal using the hyperfine couplings of 15 individual nuclear spins, and a statistical distribution for weakly coupled (bath) spins with $|\Delta| < 6$ kHz (Appendix \ref{sec:app_simulation}). c) Experimental demonstration of two-qubit gates with a detuned RF field ($N=32, \wR = 313$ Hz, $\Delta_1$ = 5, 10 kHz, $\tau= \SI{24.654}{\micro \second}$), showing lower effective Rabi frequencies compared to resonant driving (grey data, $\Delta_1 = 0$ kHz).
   \label{fig:detuned_ddrf}
  }
\end{figure}

Even though all results in this work can be generalised to other electron-nuclear spin systems, in the following we will consider in particular the $m_s=\{0,-1\}$ electron-spin subspace of the NV center in diamond ($S=1$) and its surrounding $^{13}$C nuclear spins. The main difference with other electron spin systems is how $\Delta_0$ and $\Delta_1$ depend on the hyperfine coupling. See Beukers et al. \cite{Beukers_control_2024} for experiments and simulation on an electron spin-1/2 system (the tin-vacancy center in diamond), for which the dependence of $\Delta_0 + \Delta_1$ vanishes up to second order corrections due to the perpendicular hyperfine component $A_\perp$\cite{nguyen_quantum_2019, zahedian_blueprint_2024, taminiau_universal_2014}.

All experimental results are obtained from a single NV center in a natural abundance (1.1\%) $^{13}$C diamond sample at cryogenic temperatures (4 K), with a $B_z =189.1$ mT magnetic field along the NV symmetry axis (Appendix \ref{sec:experimental_methods}). At this field, the nuclear quantisation axes are approximately parallel to the $z$-axis, and $A_\perp$ only contributes as a frequency shift (Appendix \ref{sec:sup_nv_hamiltonian}).

To verify equations \ref{eq:phase_update_rule} and \ref{eq:eff_rabi} experimentally, we perform nuclear spin spectroscopy using DDRF, by varying both the RF frequency and single-pulse phase increment $\df$ (Fig. \ref{fig:detuned_ddrf}a, similar to Fig. \ref{fig:DDRF_intro}d). First the electron spin is initialized in a superposition. A drop in measured electron coherence after application of the DDRF gate indicates interaction between the electron spin and one or more nuclear spins. 

We observe a number of traces that all follow the predicted resonance condition (Eq. \ref{eq:phase_update_rule}). The spectrum shows isolated traces indicating interactions with single nuclear spins, and a broad band-like feature corresponding to a bath of weakly coupled spins. The measured data is well-recreated by a numerical simulation modelling 15 individual spins (see Table \ref{tab:a_par}), together with a statistically distributed spin bath of many weakly coupled spins (Fig  \ref{fig:detuned_ddrf}b, see Appendix \ref{sec:app_simulation} for simulation details). 

Next, we show that the phase-increment condition (Eq. \ref{eq:phase_update_rule}), together with the single-pulse bandwidth, enables the construction of electron-nuclear gates even if the RF driving frequency is far off resonance ($\Delta_1 \gg \wR$). We perform such detuned gates on a single nuclear spin and compare them to an on-resonant gate applied to the same spin (Fig. \ref{fig:detuned_ddrf}c). All gates achieve near-unity contrast (up to some decay due to experimental noise), though the detuned version yields a reduced gate speed as predicted by Eq. \ref{eq:eff_rabi}. We further confirm the entangling nature of detuned gates by evaluating the process matrix obtained from numerical simulations (see Appendix \ref{sec:sup_optimal_phase}).

The discussion in this section shows that nuclear-spin resonance conditions are set by both the RF frequency and the single-pulse phase increment. Furthermore, quantum gates can be constructed even with significantly detuned driving frequencies by properly updating the pulse phases. This insight expands the parameter space from which gates and sensing sequences can be constructed, yielding additional possibilities for optimisation. 

\section{Weak-coupling regime ($\Delta \tau \lesssim \pi$)} \label{sec:global_window}
\begin{figure}[t]
  \includegraphics[width=1 \columnwidth]{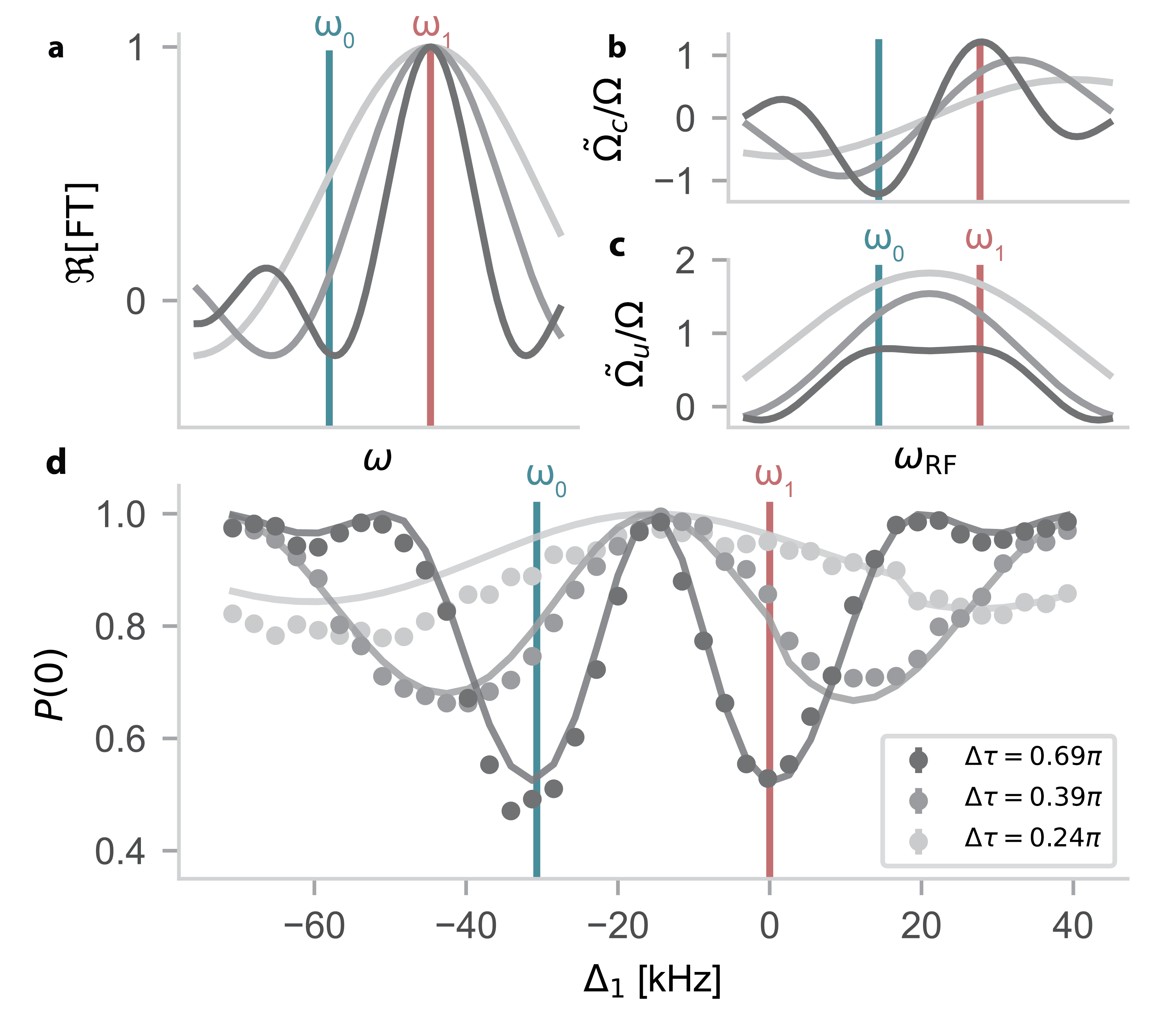}
  \caption{\emph{Weak-coupling regime.} a) Schematic showing the Fourier transform of a single RF pulse (as in Fig. \ref{fig:DDRF_intro}c) applied at $\wrf = \omega_1$, for $\Delta \tau \lesssim \pi$. As $\tau$ decreases, the pulse bandwidth increases, leading to the addressing of both the $\omega_1$ and $\omega_0$ transitions. b) The effective Rabi frequency $\tilde{\Omega}_c$ for a conditional gate (as a function of $\wrf$) is enhanced or suppressed when both transitions are addressed (Eq. \ref{eq:eff_rabi}). c) Same as in (b), but for an unconditional gate (Eq. \ref{eq:eff_rabi_uncon}).
  d) Experimental spectroscopy data as in Fig. \ref{fig:detuned_ddrf}a confirming the effective Rabi frequency for conditional gates for various $\Delta\tau$ (fixing the gate time $2N\tau\approx 1.4$ ms, $\Omega=313$ Hz). Here, $\df$ is set to track the phase increment resonance condition of a single nuclear spin ($C_0, \Delta = -30.7$ kHz, black dashed line in Fig. \ref{fig:detuned_ddrf}a). 
  Shorter $\tau$ leads to a contrast reduction and a shift of the optimal RF frequency. The analytical prediction (solid grey lines), using no free parameters, is calculated using $P(0) = 1/2\, (1 + \cos 2 \effrabi N \tau)$, with $\effrabi$ taken from (b). We attribute the model-data discrepancy to the specifics of the RF pulse envelope shape and variable RF transmission, which have not been taken into account (Appendix \ref{sec:experimental_methods})
  \label{fig:global_window}
  }
\end{figure}

When the bandwidth related to the RF pulses ($\sim 1/\tau$) is larger than, or on the same order as the hyperfine splitting $\Delta$, it is no longer valid to assume driving of only one of the nuclear spin transitions $\omega_1$ and $\omega_0$ (Fig. \ref{fig:global_window}a, Eq. \ref{H_rot}). 
In this readily encountered regime, rotations for the electron $\ket{0}$ state can cancel (or add to) the rotations for the electron $\ket{1}$ state, reducing (or enhancing) the effective rotation.

We first consider the conditional gate (Eq. \ref{eq:vcrot}), for which the $\pi$ phase shift in $\df$ inverts the RF rotation axis between subsequent pulses. In the limit of small $\Delta \tau$, $\effrabi$ is strongly attenuated (Eq. \ref{eq:eff_rabi}), proportional to $(\Delta\tau)^{-2}$ for on-resonance addressing ($\Delta_1 = 0$), and proportional to $(\Delta\tau)^{-1}$ for the optimal driving condition discussed in the next section (see Fig. \ref{fig:sensing_optimisation}b). Additionally, the extrema of $\effrabi$ shift away from the $\omega_1, \omega_0$ transitions (Fig. \ref{fig:global_window}b and Eq. \ref{eq:optimal_detuning}). This explains why the spectroscopy signals of weakly coupled spins ($\Delta \ll 1/\tau$) are suppressed and appear at detuned frequencies (broad features in Fig. \ref{fig:detuned_ddrf}a).

This analysis reveals an inherent trade-off present in DDRF gates. While a short interpulse delay $\tau$ improves the electron spin's coherence \cite{abobeih_onesecond_2018}, it also reduces the effective Rabi frequency, thereby increasing the total gate duration $T = \pi/\effrabi$ required for a $\pm \pi/2$ gate, or requiring an increase in $\Omega$.

We experimentally validate Eq. \ref{eq:eff_rabi} by driving nuclear spin $C_0$ at different RF frequencies, while updating the RF phases according to Eq. \ref{eq:phase_update_rule} (Fig. \ref{fig:global_window}d). This amounts to tracking the nuclear resonance condition apparent in Fig. \ref{fig:detuned_ddrf}a. Such a measurement directly yields the spectral signature of the reduced Rabi frequency, which is in good agreement with Eq. \ref{eq:eff_rabi}.

We repeat this measurement for different RF pulse durations $\tau$, keeping the RF amplitude ($ \wR = 313$ Hz at $\omega_1$) and the total DDRF driving time ($\approx$ 1.4 ms) fixed using $N$. For shorter $\tau$ a decrease in signal contrast and a shift of the optimal RF frequency can be observed, as predicted by Eq. \ref{eq:eff_rabi}.

The DDRF gate can also be used to perform an unconditional rotation of the nuclear spin, by leaving out the $\pi$ phase shift from $\df$ (Eq. \ref{eq:phase_update_rule}) \cite{bradley_tenqubit_2019}. The DDRF gate unitary then becomes \begin{equation}
\label{eq:uncon_U}
    V_{\mathrm{ROT}} = \ide \otimes R_x (N \Omega_u\tau)\,, 
\end{equation}with effective (unconditional) Rabi frequency:
\begin{equation} \label{eq:eff_rabi_uncon}
    \tilde{\Omega}_u=\wR \left(  \mathrm{sinc}(\Delta_1 \, \tau) + \mathrm{sinc}(\Delta_0 \, \tau) \right) \,.
\end{equation}
In contrast to the conditional case, $\tilde{\Omega}_u$ is enhanced at small $\Delta\tau$ (Fig. \ref{fig:global_window}c), as without the extra $\pi$ phase shift the RF rotations build up constructively. In the limit $\Delta \ll 1/(2\tau)$ the effective Rabi frequency $\Omega_u$ approaches $2\Omega$, identical to constant RF driving of a nuclear-spin transition while keeping the electron spin in an eigenstate.

We identify three approaches for mitigating the reduced Rabi frequency for conditional gates. First, for a fixed gate length $2N\tau$, the number of decoupling pulses $N$ can be traded for RF pulse length ($\approx 2\tau$), shrinking the pulse bandwidth to avoid driving both transitions.
This comes at the cost of decreasing effectiveness of the electron decoupling, as longer interpulse delays protect less effectively against low-frequency noise \cite{abobeih_onesecond_2018}. Second, given a certain pulse length, the RF frequency can be detuned to maximise Eq. \ref{eq:eff_rabi}. Third, one could increase the physical RF amplitude to compensate for the reduction in Rabi frequency, although this poses experimental challenges, and our model validity is constrained to $\Omega\tau \ll 1$. The next sections explore these approaches in more detail, in the context of nuclear-spin sensing and multi-qubit control.

\section{Optimal sensitivity for sensing a single nuclear spin} \label{sec:sensing_optimisation}
To highlight the practical significance of the presented insights, we demonstrate how to optimise the DDRF sequence for sensing a single, weakly coupled nuclear spin (with hyperfine coupling $\Delta$). For example, this nuclear spin could be a single proton or $^{13}$C spin, potentially outside of the host crystal \cite{schwartz_blueprint_2019}. The goal is to minimise the (single-spin) sensitivity, defined as \cite{degen_quantum_2017}(Appendix \ref{sec:sup_sensitivity}):
\begin{equation} \label{eq:sensitivity}
    \sens = \frac{2\pi e^{\chi(N, t)}}{\effrabi_{\mathrm{max}}(\Delta,N,t) \sqrt{t}} \, ,
\end{equation}
where $\chi(N,t)$ is the sensor decoherence function (here taken from the experimental observations of Ref. \cite{abobeih_onesecond_2018}), $\effrabimax$ is the maximum attainable effective Rabi frequency (given $\Delta$ and $N$), and $t=2N\tau$ is the single-experiment sensing time \cite{degen_quantum_2017}. For simplicity, we assume unity readout contrast and zero sensor overhead (Appendix \ref{sec:sup_sensitivity}).

The expression for $\sens$ in Eq. \ref{eq:sensitivity} conveys the minimum number of nuclear spins required that together yield a detectable signal in 1 s of integration time. Evidently, to achieve single-spin sensitivity ($\sens<1$), the \emph{effective} Rabi frequency $\effrabi$, which sets the effective coupling to the signal, should be as large as possible, while retaining sufficient electron coherence ($e^{-\chi(N, t)}$). The choice of $N$ presents us with an inherent trade-off between these two factors. Generally, larger $N$ (shorter $\tau$-values) increase the electron coherence, but decrease $\effrabimax$ (see Sec. \ref{sec:global_window}). However, optimising over the large parameter space is challenging.

\begin{figure}
  \includegraphics[width=1 \columnwidth]{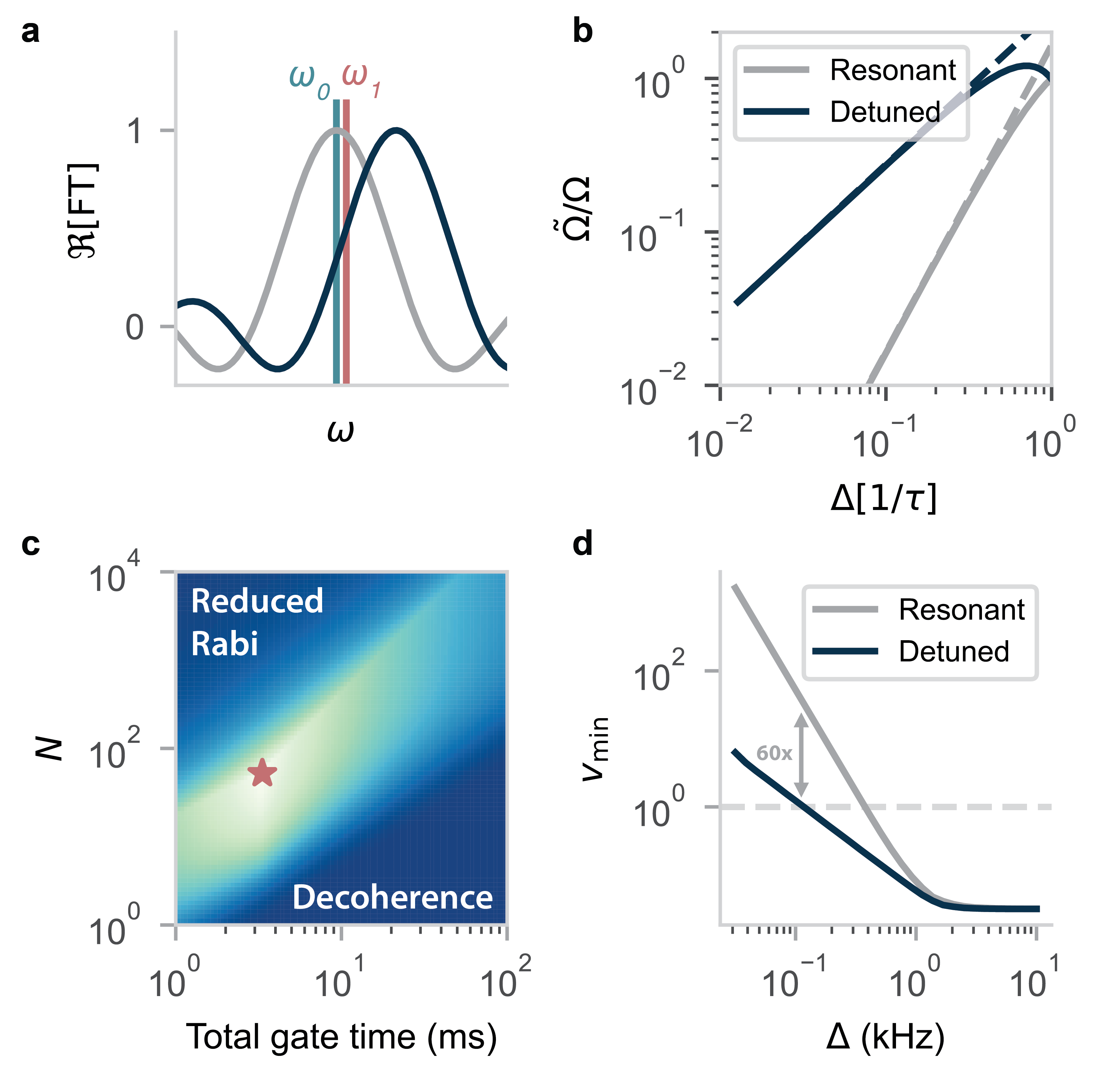}
  \caption{\emph{Optimised detuned sensing.} a) Schematic illustrating maximum DDRF contrast if the $\omega_0$ and $\omega_1$ transitions are on the slope of the RF pulse envelope. b) Numerical calculation of the Rabi frequency suppression as a function of $\Delta \tau$ for $\Delta < \tau$. When driving on resonance, $\effrabi \propto (\Delta \tau)^2$, however by using the optimal detuning (Eq. \ref{eq:optimal_detuning}) $\effrabi \propto \Delta \tau$ (scaling indicated by dashed lines). The effective Rabi frequency can thus be much improved when purposefully driving off-resonance. c) Inverse sensitivity ($v_{\mathrm{min}}^{-1}$), varying the number of pulses $N$ and total sequence time $t$. Short inter-pulse spacing $\tau$ (corresponding to short $t$ and high $N$) leads to the driving of both transitions, while long $\tau$ (i.e. using low $N$) leads to electron decoherence. The red star denotes the optimal parameters. d) Detuned sensing achieves orders of magnitude higher sensitivity, still allowing for single-spin sensitivity (grey dotted line) for small $\Delta$.  
  \label{fig:sensing_optimisation}
  }
\end{figure}

We first reduce the parameter space size by calculating the RF detuning that maximises $\effrabi$. For $\Delta \gtrsim 2\pi/\tau$, the RF driving when the electron in in the $\ket{0}$ state can be neglected and the optimal effective Rabi frequency is always attained when driving on resonance ($\Delta_1 = 0$). However, when $\Delta \lesssim 2\pi/\tau$,  significant enhancement is possible by detuning the RF frequency. We find that the optimum setting for $\Delta_1$ is (approximately) given by (Appendix \ref{sec:sup_optimal_rf_frequency}):

\begin{equation}\label{eq:optimal_detuning}
  \Delta_1 = \begin{cases}
    - \frac{\ws}{\tau} + \Delta / 2, & \text{if $\tau \lesssim 2\pi/\abs{\Delta}$}.\\
    0, & \text{otherwise}.
  \end{cases}
\end{equation}
where $\ws \approx 2.082$ is the first root of the second derivative of the sinc function. Conceptually, this condition is satisfied when the detuning is such that the gradient of the RF pulse envelope is maximal in between the $\omega_0$ and $\omega_1$ transitions (Fig. \ref{fig:sensing_optimisation}a). While for $\Delta_1=0$, $\effrabi \propto (\Delta\tau)^2$, using the optimum $\Delta_1$ changes the scaling to $\effrabi \propto \Delta\tau$ (Fig. \ref{fig:sensing_optimisation}b). We verify that Eq. \ref{eq:optimal_detuning} maximises Eq. \ref{eq:eff_rabi} in the small $\Delta$ regime (Appendix \ref{sec:sup_optimal_rf_frequency}).

Next, we evaluate the optimal RF amplitude. For the situation of a very weakly coupled spin, the reduced Rabi frequency can be partially compensated for by increasing the physical RF amplitude $\wR$. However, simply setting the RF amplitude to the inverse of $\effrabi/\wR$ leads to unrealistically high values when $\tau \ll 2\pi/\Delta$. Moreover, our model of the effective Rabi frequency strictly only holds for $\wR \tau \ll 1$. Therefore, in the current analysis, we set an upper bound for the RF amplitude: $\wR < 1/(2\tau)$. Additionally, we limit the maximum value to $\wR < \SI{10}{\kilo \hertz}$, as higher Rabi frequencies are typically challenging to reach without specialised RF transmitters, especially at cryogenic temperatures \cite{herb_broadband_2020,yudilevich_coherent_2023}. 

Then, to find the optimal sensing parameters, we evaluate (the inverse of) Eq. \ref{eq:sensitivity}, sweeping the number of pulses $N$ and total sequence time $t = 2N\tau$, while continuously updating the RF detuning and RF amplitude to maximise $\effrabi$ (Fig. \ref{fig:sensing_optimisation}c and Appendix \ref{sec:sup_sensing_optimisation_procedure}). As expected, there exists an optimal regime that balances the expected reduction in Rabi frequency and electron decoherence (bright sliver in Fig. \ref{fig:sensing_optimisation}c). We compare the sensitivity of the (conventional) resonant gate with the detuned protocol by extracting the optimal value for various $\Delta$ (Fig. \ref{fig:sensing_optimisation}d) and find that the latter outperforms the former for small $\Delta$. A divergence between the two can be observed at $\Delta/(2\pi) \sim 1/T_2 \approx 1$ kHz, with the detuned protocol still achieving single-spin sensitivity at a mere $115$ Hz hyperfine coupling, a performance enhancement by a factor 60. Conversely, a statistically polarised ensemble of 100 $^{13}$C spins would be detectable from a distance of $\sim 26$ nm, compared to $\sim 6$ nm for the resonant protocol (assuming the spin coherence of Ref. \cite{abobeih_onesecond_2018} continues to hold). 

\section{Quantum gate selectivity}\label{sec:gate_optimisation}

\begin{figure*}
    \centering
    \includegraphics[width=\textwidth]{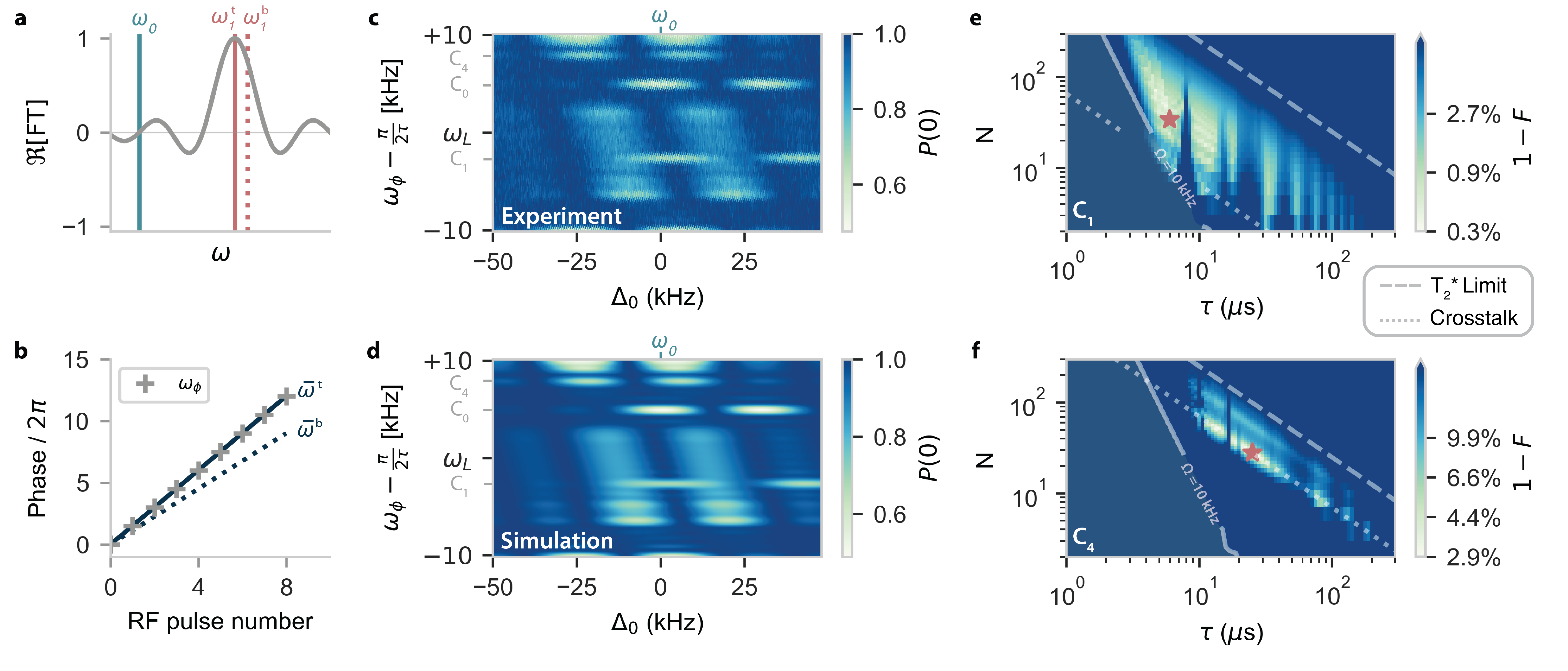}
    \caption{\emph{Quantum gate selectivity.} a) Schematic as in Fig. \ref{fig:DDRF_intro}c. Due to their high bandwidth, single RF pulses may drive transitions of bystander spins ($\omega_1^{b}$, dotted line), potentially inducing crosstalk (Eq. \ref{eq:rf_sel}). b) Schematic showing the phase build-up for the target spin (solid blue line) and the bystander spin (dotted blue line). The target spin can be selectively addressed by implementing the appropriate phase increment (grey crosses, Eq. \ref{eq:phase_update_rule}, for clarity plotted without the extra $\pi$ increments). c) Experimental spectroscopy data as in Fig. \ref{fig:detuned_ddrf}a, with different $\tau=\SI{24.654}{\micro \second}$. Plotting the data as a function of the phase-increment frequency $\wf$ (Eq. \ref{eq:phase_freq}) and $\Delta_0$ highlights the different selectivity mechanisms. 
    Nuclear spins appear at $\wf - \pi/(2\tau) = \wmean$ (linewidth $\sim (2N\tau)^{-1}$), modulated along the $\Delta_0$ ($\propto \wrf$) axis by their effective Rabi frequency ($C_0$, $C_1$ and $C_4$ resonances annotated).
    The bandwidth of the $y$-axis (here $\sim\SI{20}{\kilo \hertz}$) is set by the $\pi/\tau$ periodicity of $\Df$, resulting in aliases that can cause spins to overlap. Jagged artifacts are due to the limited resolution of the data (taken as a function of $\df$). d) Simulation of (c) (Appendix \ref{sec:app_simulation}). e) Numerical simulation of the maximum attainable DDRF gate fidelity for nuclear spin $C_1$ in a 6-spin register (including $C_0, C_1, C_4, C_6, C_8$, see Table 
    \ref{tab:a_par}). The fidelity is calculated for the full 6-qubit unitary (red star denotes optimal parameters). The dotted line indicates $2N\tau=131\,\mu$s, the minimum gate time that satisfies the selectivity bound (Eq. \ref{eq:phase_sel_nv}). The dashed line indicates $2N\tau=5\,$ms, corresponding to the limit on gate time imposed by nuclear-spin $T_2^*$. The color bar has logarithmic spacing. f) Simulation for $C_4$, a more weakly coupled and spectrally crowded spin, resulting in lower fidelities. Here the dotted line indicates $2N\tau=1.4\,$ms.}

    \label{fig:selectivity}
\end{figure*}

Finally, we consider the use of the DDRF gate sequence for qubit control \cite{abobeih_faulttolerant_2022, bartling_universal_2024,bradley_tenqubit_2019}. The challenge is to realise a high-fidelity (two-qubit) gate on a selected nuclear spin, while avoiding crosstalk to other spins. First, we consider the selectivity of the gates starting from the above results (Eqs. 1-5). In the next section, we simulate a realistic spin register and identify the parameter regime(s) in which high fidelity gates are possible.

We identify two selectivity mechanisms for the DDRF gate: selective driving due to the limited bandwidth of a single RF pulse (Fig. \ref{fig:selectivity}a), and the constructive build-up of small rotations due to the phase-increment condition being met for a specific nuclear spin (Fig. \ref{fig:selectivity}b). Selective control can be achieved through either mechanism, or through a combination thereof (see e.g. Fig. \ref{fig:selectivity}c and d).  To quantitatively study these mechanisms, we consider a target nuclear spin qubit $t$, and a second, bystander, nuclear spin $b$ for which crosstalk is to be avoided. 

For the selectivity stemming from the individual RF pulses, Eq. \ref{eq:eff_rabi} can directly be used to yield the smallest difference in nuclear-spin transitions for which $\effrabi_b=0$:
\begin{equation}
\label{eq:rf_sel}
    \left|\Delta_{0/1}^{t} - \Delta_{0/1}^{b}\right| \gtrsim \frac{\pi}{\tau} \,.
\end{equation}
The superscripts $t$ and $b$ are used to denote the target and bystander spins, respectively, and the '0/1' subscript implies that this condition must hold for any combination of the $\omega_0$ and $\omega_1$ transitions. Evidently, the selectivity stemming from the RF pulses is limited by their bandwidth ($\sim 1/\tau$), and the proximity of the nuclear spin transitions. For the NV electron spin-1 system considered here, the minimum difference between nuclear-spin hyperfine couplings required for achieving selectivity within a single RF pulse is
\begin{equation}\label{eq:freq_sel_nv}
    \left|A_{\parallel}^{t} - A_{\parallel}^{b} \right| \gtrsim \frac{\pi}{\tau} \,.
\end{equation}

Next, to describe the selectivity due to the phase increments, it is instructive to realise that the phase increments effectively modulate the bare RF frequency (also known as `phase ramping' \cite{vandersypen_nmr_2005a}), so that we can define a phase-increment frequency:
\begin{equation}\label{eq:phase_freq}
    \omega_\phi = \wrf + \frac{\df}{2\tau}\,.
\end{equation}
We can then rewrite the phase-increment resonance condition (Eq. \ref{eq:phase_update_rule}) as: 
\begin{equation}\label{eq:wphi}
    \omega_\phi = \bar{\omega} + \frac{(2k+1)\pi}{2\tau},
\end{equation}
with $k \in \mathbb{N}$. Here, selectivity arises due to the difference in the mean frequencies $\bar{\omega}$ of the target and bystander spins, which has to be large enough for selective control.

A lower bound on the mean frequency difference can be attained by considering the Fourier-limited frequency resolution of the phase ramp (determined by its length $\sim 1/(2N\tau)$):
\begin{equation}
\label{eq:phase_sel}
    \left|
    \left(\wmean^{\mathrm{t}}\mod \frac{\pi}{\tau}\right)-\left(\wmean^{\mathrm{b}}\mod \frac{\pi}{\tau}\right)
    \right|
    \gtrsim \frac{\pi}{N \tau}\,, 
\end{equation}
where the modulo stems from the $i/\tau$ periodicity in Eq. \ref{eq:wphi}. Although Eq. \ref{eq:phase_sel} strictly speaking constitutes an upper bound to the selectivity, it allows for a potential selectivity enhancement by a factor $N$ compared to Eq. \ref{eq:rf_sel}. Whether such an enhancement is possible in practice depends on the the spectrum of $\wmean$ for the electron-nuclear spin system of interest. For example, for a spin-1/2 defect spin, the left-hand side of Eq. \ref{eq:phase_sel} vanishes, up to second order corrections due to the perpendicular hyperfine component $A_\perp$ \cite{Beukers_control_2024,nguyen_quantum_2019, zahedian_blueprint_2024}.

For the spin-1 system considered here, an exact bound for the selectivity can be derived, under the assumption of negligible driving in the electron $\ket{0}$ state (Appendix \ref{sec:app_gate_sel}). In particular, fixing $\Omega N \tau = \pi/2$ to create a fully entangling gate, the condition for a selective gate on the target spin is given by:
\begin{equation}\label{eq:phase_sel_nv}
    \left|\left(\frac{A_{\parallel}^{t}}{2}\mod \frac{\pi}{\tau}\right) - \left(\frac{A_{\parallel}^{b}}{2}\mod \frac{\pi}{\tau}\right) \right| \gtrsim \frac{\sqrt{15}\pi}{4 N\tau}\,.
\end{equation}

To illustrate the selectivity mechanisms, we again perform DDRF spectroscopy (similar to Fig. \ref{fig:detuned_ddrf}a), but instead of $\df$, we now plot the data as a function of $\omega_\phi - \pi/(2\tau)$ (Fig. \ref{fig:selectivity}c, d). This is a useful quantity as it is independent of the RF frequency, and directly relates to a spin's mean frequency. Spins appear at their $\bar{\omega}$ frequency along the $\omega_\phi$ axis ($\mathrm{mod}\ \pi/\tau$), with their signal intensity modulated by the effective Rabi frequency (Eq. \ref{eq:eff_rabi}), which varies with $\Delta_0$ ($x$-axis). The signal from the spin bath (slanted band-like features) is pushed away from $\Delta_0=0$ due to the form of Eq. \ref{eq:eff_rabi} in the weak-coupling regime (section \ref{sec:global_window}).

The widths of the spin response along both axes partly determine if the spin can be selectively controlled or overlaps with other spins (crosstalk), as given in equations \ref{eq:freq_sel_nv} and \ref{eq:phase_sel_nv}.
The $\frac{\pi}{\tau}$ periodicity of $\omega_\phi$ creates opportunities for unexpected crosstalk to occur. For example, both the $\omega_0$ and $\omega_1$ transition of nuclear spin $C_1$ ($\Delta = \SI{-45.9}{\kilo\hertz}$, Table \ref{tab:a_par}), somewhat overlap with sidelobes of the spin bath, limiting the expected gate fidelity of that spin for these gate parameters. In contrast, the transitions of spin $C_0$ ($\Delta = \SI{-30.7}{\kilo \hertz}$, Table \ref{tab:a_par}) is not affected by such crosstalk, due to the particular value of $\tau$ used here. Note that for electron spin-1/2 systems, all nuclear spins will appear at approximately $\omega_\phi -\pi/(2\tau)= \omega_\mathrm{L}$ (up to second order corrections due to $A_\perp$), so that selective control depends more on whether spins can be resolved along the $\Delta_0$-axis \cite{Beukers_control_2024}.

\section{A multi-qubit nuclear-spin register}
\label{sec:multi_spin_register}
We apply the insights from the previous section to investigate the boundaries of the gate parameter space that allows for high-fidelity control in a multi-qubit nuclear-spin system. Considering the 15 identified spins near this NV center (Table \ref{tab:a_par}), we select a register of 5 nuclear spins that are most isolated in frequency space (Appendix \ref{sec:register_optimisation}). 

For a register with $M$ nuclear spins, the ideal operation is given by:
\begin{align} \label{eq:target_operation}
    \Uideal = \mathrm{CR}_{\mathrm{x}}(\pm \pi/2) \otimes \ide^{\otimes (M-1)}\, ,
\end{align}
where the controlled rotation acts on the electron and target spin subspace. %

In addition to the unitary evolution of the 6-qubit register dictated by $H_{\mathrm{RF}}$ (Eq. \ref{H_rot}), we include three additional contributions to the infidelity (Appendix \ref{sec:register_optimisation}): (i) the electron-spin $T_2$ dephasing time under dynamical decoupling (phenomenological, taken from Abobeih et al. \cite{abobeih_onesecond_2018}). (ii) Electron-spin dephasing due to the direct interaction (Eq. \ref{H_rot}) with the characterised nuclear spins outside the register, and the nuclear spin bath (in a mixed state). (iii) $T_2^\star$ dephasing ($\approx \SI{10}{\milli \second}$\cite{bradley_tenqubit_2019}) of the nuclear spins in the register, simulated as quasi-static magnetic field noise.
As is commonly done in experimental settings, we introduce a post-gate echo pulse on the register spins to partially mitigate this dephasing \cite{abobeih_faulttolerant_2022, bradley_tenqubit_2019}. 
Additionally, we restrict the RF amplitude to a maximum value of $\Omega = \SI{10}{\kilo \hertz}$ (as in section \ref{sec:sensing_optimisation}).

To identify optimal gate parameters, we vary the interpulse delay $\tau$ and the number of pulses $N$, while ensuring the RF amplitude $\Omega$ is set to create the desired $\pm\pi/2$ rotation (compensating for the reduction in effective Rabi frequency using Eq. \ref{eq:eff_rabi}). We set $\wrf$ to the target spin's $\omega_1$ transition and calculate $\df$ accordingly (Eq. \ref{eq:phase_update_rule}), applying a second-order correction due to the AC-Stark shift (Appendix \ref{sec:sup_optimal_phase}). 

We obtain (6-qubit) average gate fidelity maps \cite{nielsen_simple_2002}, such as presented in Fig. \ref{fig:selectivity}e and f, for target nuclear spins $C_1$ and $C_4$, respectively (Table \ref{tab:a_par}). We identify the following bounds on the available parameter space for high-quality gates. First, the maximum gate time is limited by nuclear- and electron-spin decoherence, the latter of which depends on $N$ and $\tau$. A minimum gate time is dictated by the degree of spectral crowding of the target spin, as predicted by Eq. \ref{eq:phase_sel}. Contrary to $C_1$, $C_4$ is spectrally close to another spin ($C_3$, $\sim \SI{1.4}{\kilo \hertz}$), so that high-fidelity gates are only found for larger gate times ($2N\tau \gtrsim \SI{1.4}{\milli \second}$). In particular, through the definition of the selectivity (Appendix \ref{sec:app_gate_sel}), equalising Eq. \ref{eq:phase_sel_nv} (dotted line in Fig. \ref{fig:selectivity}f) ensures zero crosstalk with the nearest bystander spin ($C_3$).

Furthermore, even though the reduction in effective Rabi frequency $\effrabi$ can in principle be fully compensated for, when $\Omega \sim \Delta$, the assumptions underlying Eq. \ref{eq:eff_rabi} break down (\ref{sec:sup_eff_rabi_derivation}), and our prediction for the optimal gate parameters no longer produces high-fidelity gates (top left corner of Figs. \ref{fig:selectivity}e and f). For spins with smaller hyperfine couplings (e.g. $C_4$), this effect is more detrimental to the maximum attainable fidelity. Finally, the remaining parameter space is interspersed with sharp drops in fidelity at $\tau$-values for which crosstalk occurs with other individual spins, or with the nuclear spin bath. The combination of these effects results in maximimum 6-qubit gate fidelities of $F = 99.7\%$ and $F=97.1\%$ for target spins $C_1$ and $C_4$, respectively. See Appendix \ref{sec:register_optimisation} for the other register spins, and breakdowns of the different infidelity contributions.

Note that here we did not optimize $\wrf$, which could further reduce crosstalk and realize improved effective Rabi frequencies (see section \ref{sec:sensing_optimisation}).

\section{Conclusions} \label{sec:conclusions}

In conclusion, we presented an improved, and generalised framework for electron-nuclear DDRF gates. Our model reveals that these gates can be deconstructed into two independent components: (1) the driving induced by individual RF pulses (Eq. \ref{eq:eff_rabi}) and (2) the RF phase increments that enable constructive rotational build-up throughout the sequence (Eq. \ref{eq:phase_update_rule}). Considering these components independently allows for increased versatility in gate optimisation for sensing and quantum control (e.g. by detuning the RF frequency), but also reveals inherent limitations in gate speed and selectivity.

A first, general, insight is that the \emph{effective} Rabi frequency for short interpulse delays ($\Delta \tau \lesssim \pi$) can be strongly suppressed for conditional gates and enhanced for unconditional gates. This reveals an inherent trade-off between protecting electron spin coherence by faster decoupling, and retaining nuclear spin selectivity and gate efficiency. This trade-off has important implications for the sensing and control of nuclear spins, including in typical physical systems, such as for an electron spin in a dilute nuclear spin bath \cite{vandestolpe_mapping_2024, cujia_parallel_2022, abobeih_faulttolerant_2022,babin_fabrication_2022,nguyen_quantum_2019}. The presented detuned sensing scheme partially compensates for this reduction in Rabi frequency, making DDRF a promising alternative to conventional dynamical-decoupling spectroscopy in the context of nano-NMR \cite{taminiau_detection_2012,staudacher_nuclear_2013,mamin_nanoscale_2013}. Future work might extend this principle to high-fidelity quantum gates. 

A second key insight is that quantum gate selectivity stems both from direct RF driving of the spin transition frequencies, as well as the targeting of the mean spin evolution frequency by the phase increments. Importantly, for systems in which the mean nuclear-spin frequencies are first-order degenerate (e.g. electronic spin-1/2 systems), the RF pulses can still provide a selectivity mechanism, albeit limited by their bandwidth (see also Beukers et al.\cite{Beukers_control_2024}). A possible improvement may consist of creating intermediate evolution periods with a nuclear-spin frequency that is dependent on the electron-spin state, for example by temporarily swapping the electron-spin state to a memory qubit \cite{Beukers_control_2024}.

These results provide new opportunities for the optimisation of quantum gate fidelities for quantum information processing and quantum network applications \cite{bradley_robust_2022, pompili_realization_2021,abobeih_faulttolerant_2022,debone_thresholds_2024a, bartling_universal_2024}. They are applicable to a large variety systems, such as various spin defects in diamond \cite{Beukers_control_2024, nguyen_quantum_2019}, silicon \cite{higginbottom_optical_2022} and silicon-carbide \cite{babin_fabrication_2022, bourassa_entanglement_2020}, and might also be transferable to other platforms such as quantum dots \cite{hensen_silicon_2020} and rare-earth ions \cite{ruskuc_nuclear_2022}.

\section*{Data and code availability}
All data underlying the study are available on the open 4TU data server \cite{ddrf_data}. Code used for performing the numerical simulations and for operating the experiments is available on request.

\section*{Acknowledgements}
We thank H. P. Bartling and C. E. Bradley for useful discussions. We gratefully acknowledge support from the joint research program “Modular quantum computers” by Fujitsu Limited and Delft University of Technology, co-funded by the Netherlands Enterprise Agency under project number PPS2007. We acknowledge financial support from the Quantum Internet Alliance through the Horizon Europe program (grant agreement No. 101080128). This project has received funding from the European Research Council (ERC) under the European Union’s Horizon 2020 research and innovation programme (grant agreement No. 852410). This work was supported by the Dutch National Growth Fund (NGF), as part of the Quantum Delta NL programme. This work is part of the research programme NWA-ORC with project number NWA.1160.18.208, which is (partly) financed by the Dutch Research Council (NWO). This project has received funding from the European Union’s Horizon Europe research and innovation program under grant agreement No
 101135699.

\appendix

\section{EXPERIMENTAL METHODS}
\counterwithin{figure}{section}

\label{sec:experimental_methods}

All experiments are conducted on a naturally occurring NV center using a custom-built cryogenic confocal microscopy setup (4K, Montana Cryostation). The diamond sample, which has a natural abundance of 1.1\% $^{13}$C, was homoepitaxially grown via chemical vapor deposition (Element Six) and cleaved along the $\langle 111 \rangle$ crystal direction.

A solid immersion lens (SIL) is milled around the NV center to improve photon collection efficiency\cite{robledo_highfidelity_2011}. A gold stripline is deposited near the edge of the SIL for the application of microwave (MW) and radio-frequency (RF) pulses. Typical nuclear Rabi frequencies can reach up to $\sim \SI{1.6}{\kilo \hertz}$, above which sample heating starts to affect the NV readout. MW and RF signals are generated by a ZI HDAWG Arbitrary Waveform Generator, in combination with a MW mixer and separate RF and MW amplifiers. An external magnetic field of $B_z = \SI{189.1}{\milli \tesla}$ is applied along the NV-symmetry axis, using a permanent neodymium magnet mounted to the back of the cryostat cold finger. An external permanent magnet is used for fine alignment of the magnetic field, the small remaining perpendicular magnetic field components are neglected here.

The NV spin state is initialized via spin-pumping and read out in a single shot through spin-selective resonant excitation, with fidelities $F_0 = 0.930(3)$ ($F_1 = 0.995(1)$) for the $m_s = 0$ ($m_s = -1$) state, respectively, resulting in an average fidelity of $F_{\mathrm{avg}}=0.963(3) $. Reported data is corrected for these numbers to obtain a best estimate of the electronic spin state. 
We drive the electronic $m_s = 0 \leftrightarrow m_s = -1$  spin transition at 2.425 GHz with Hermite-shaped pulses.

In this work, XY-8 type sequences are used for dynamical decoupling during the DDRF gate, to minimize the effects of pulse errors\cite{abobeih_onesecond_2018}. 
The length of RF pulses in the DDRF sequences in this work are set to an integer number of periods of the RF radiation, to prevent the NV electron spin from picking up extra phase. Each RF pulse has a $\sin^2(t)$ roll-on and roll-off to prevent signal ringing in the RF signal chain, with a roll-duration of two RF periods. Due to this pulse shaping, the two RF pulses of length $\tau$ in the DDRF sequence create a smaller combined rotation than the RF pulses of length $2\tau$. To correct for this, the amplitude of the single-$\tau$ pulses is multiplied by the ratio of the integrals of the $2\tau$ pulse and the two single-$\tau$ pulses.

DDRF spectroscopy (Figs. \ref{fig:detuned_ddrf},\ref{fig:global_window} and \ref{fig:selectivity}c) is performed by (i) preparing the electron spin in the $m_s=0$ state; (ii) applying a $\pi/2$ MW pulse on the electron spin to prepare the $\ket{+}$ state; (iii) Performing the DDRF sequence; (iv) applying a $-\pi/2$ MW pulse (same axis as the initial $-\pi/2$ pulse); (v) reading out the electron spin state. The experiment is repeated to estimate $P(0)$, the probability to find the electron in the $m_s=0$ state. $P(0)$ corresponds to the remaining electron spin coherence after the DDRF sequence.

\section{NV HAMILTONIAN} \label{sec:sup_nv_hamiltonian}
The NV center is described by a spin-1 electron coupled to a spin-1 nitrogen spin, with additional couplings to spin-1/2 $^{13}\mathrm{C}$ spins. This work is limited to the $m_s = {0, -1}$ subspace of the electron spin. The interaction with the spin-1 nitrogen spin is neglected, as it is initialized in the $m_N=0$ state and the decoupling sequences effectively cancel the interaction between electron and nitrogen spin. Considering one $^{13}\mathrm{C}$ spin, the Hamiltonian of the system is, in the interaction picture and after the rotating-wave and secular approximation, given by:
\begin{multline} \label{eq:full_hamiltonian}
    H = \omega_0 I_z + A_\parallel S_z I_z + A_\perp S_z I_x \\ + 2 \Omega \cos (\wrf t + \phi) I_x, 
\end{multline}
where $A_{\parallel} = A_{zz}$ and $A_{\perp}= \sqrt{A_{zx}^2 + A_{zy}^2}$ are the parallel and perpendicular components of the NV-nuclear hyperfine tensor, $S_z$, $I_z$ and $I_x$ are the electron and nuclear (pseudo) spin-1/2 operators, $\Omega$ is the physical RF amplitude, $\wrf$ is the RF frequency, and $\phi$ is some phase offset of the RF field. Note that, in general, the axis of $A_\perp$ and the axis along which the RF-radiation is applied are not the same. In the present work, for simplicity, the effect of $A_\perp$ is neglected \cite{bradley_tenqubit_2019}, further motivated by the high magnetic field (189.1 mT) at which experiments were performed. At such high fields, the tilt of the nuclear quantisation axis is small ($< \SI{1}{\degree}$ for $A_\perp \sim \SI{30}{\kilo \hertz}$). We do take into account the changed nuclear precession frequency due to $A_\perp$, given by: $\omega_1 = \sqrt{A_\perp^2 + (\omega_0 - A_\parallel)^2}$. In the rotating frame at the RF frequency (and again making the rotating wave approximation) Eq. \ref{eq:full_hamiltonian} simplifies to:
\begin{multline}
     H_{\mathrm{RF}} = \ket{0}\bra{0} \otimes \Delta_0 I_z 
    + \ket{1}\bra{1} \otimes \Delta_1 I_z \\
    + \ide \otimes \wR (\cos \phi I_x + \sin \phi I_y),        
\end{multline}

\section{SIMULATION OF DDRF SPECTROSCOPY}\label{sec:app_simulation}
Spectroscopy experiments (Figs. \ref{fig:detuned_ddrf}b, \ref{fig:selectivity}d) have been simulated assuming the presence of 15 individual nuclear spins (Table \ref{tab:a_par}) and a bath of weakly coupled spins. %

For an individual nuclear spin $c$, the unitary operation $U$ of an $N=2$ DDRF unit-cell was calculated starting from $H_{\mathrm{RF}}$ (Eq. \ref{H_rot}). For starting electron state $\ket{0}$ ($\ket{1}$), the rotation axis $\mathbf{\hat{n}}_{0,c}$ ($\mathbf{\hat{n}}_{1,c}$) and angle $\theta_{0,c}(=\theta_{1,c})$ of the nuclear spin rotation was determined. If the electron spin starts in the $\ket{x}$ state, the $\left<\sigma_x\right>_c$ expectation value after a DDRF gate with $N$ pulses is given by \cite{taminiau_detection_2012}
\begin{align}\label{eq:sim_sigma}
    \left<\sigma_x\right>_c = 1 - (1 - \mathbf{\hat{n}}_{0,c} \cdot \mathbf{\hat{n}}_{1,c})\sin^2 \frac{N\theta_c}{2}\, .
\end{align} 
The total signal from the individual spins is given by the product of the expectation values:
\begin{align}
    \left<\sigma_x\right> = \prod_c \left<\sigma_x\right>_c
\end{align}

For the nuclear spin bath, a mean density of parallel hyperfine shifts $\Delta$ is used \cite{vandestolpe_mapping_2024}
\begin{align}
    \rho(\Delta) = \frac{\pi^2 \alpha \rho_{^{13}\mathrm{C}}}{\Delta^2},
\end{align}
where $\alpha = \hbar\mu_0\gamma_e\gamma_c/4\pi$, and $\rho_{^{13}\mathrm{C}} = n_{^{13}} \rho_{C} = 1.950\mathrm{nm}^{-3}$ is the density of $^{13}\mathrm{C}$ in natural abundance diamond, where $n_{^{13}}$ is the relative abundance of $^{13}\mathrm{C}$ atoms in the environment ($1.109\%$), and $\rho_{C}$ is the density of C atoms in diamond. 

Instead of calculating the bath signal for a random sample of individual spins\cite{abobeih_onesecond_2018}, we calculate the signal for a sufficient number of bins (300 in this work) of $\Delta$ of width $d\Delta$. The expected number of spins in such a bin is given by $\rho(\Delta)d\Delta$. The spectroscopy signals for each bin $\left<\sigma_x(\Delta)\right>$ are calculated with Eq. \ref{eq:sim_sigma} and combined to yield the total signal of the spin bath:\begin{align} \label{eq:nuclear_spin_bath}
    \left<\sigma_x\right>_{\mathrm{bath}} = \prod_{\Delta=-\Delta_{\mathrm{lim}}}^{\Delta_{\mathrm{lim}}}\left<\sigma_x(\Delta)\right>^{\rho(\Delta)d\Delta}
\end{align}
where $\Delta_{\mathrm{lim}}$ defines the maximum coupling strength of spins that are still considered part of the spin bath, in this work $\Delta_{\mathrm{lim}} = 2\pi\times 6$ kHz. Note that this approach breaks down when the signal due to a single spin with $\Delta$ becomes large. The total signal is given by the product of the signals of individual spins and the spin bath.

The Rabi frequency $\Omega$ was determined from the waveform amplitude of the RF pulses and an experimentally determined conversion factor from waveform amplitude to $\Omega$. The pulse length $\tau_{\mathrm{rf}}$ is adjusted by half the length of the pulse roll-on time (Appendix \ref{sec:experimental_methods}) to approximately account for the smaller effective pulse amplitude, and the dead-time around the MW pulses is subtracted. The RF Hamiltonian is applied for a duration of $\tau_{\mathrm{rf}}$, and during the dead-time we set $\Omega=0$. The simulation could be made more accurate by considering a time-dependent $\Omega$, matching the pulse shape, at the cost of computation time.

From the spectroscopy signal, we qualitatively identify 15 individual nuclear spins that can be distinguished from the spin bath (listed in Table \ref{tab:a_par}). 
\begin{table}[h]
\centering
\begin{tabular}{|c|c|c|c|c|c|}
\hline
Index & $\Delta$ \text{(Hz)} & Index & $\Delta$ \text{(Hz)} & Index & $\Delta$ \text{(Hz)} \\
\hline
$C_0^*$  & -30693  & $C_5$  & -12570  & $C_{10}$ & -9500   \\
$C_1^*$  & -45870  & $C_6^*$  &  15744  & $C_{11}$ & -9000   \\
$C_2$  &  20000  & $C_7$  & -10020  & $C_{12}$ & -13060  \\
$C_3$  &  19900  & $C_8^*$  & -11160  & $C_{13}$ & -6193   \\
$C_4^*$  &  18500  & $C_9$  &  -7660  & $C_{14}$ & -7200   \\
\hline
\end{tabular}
\caption{Characterised nuclear spin hyperfine shifts. Spins marked with * form the register considered in Section \ref{sec:gate_optimisation}.}
\label{tab:a_par}
\end{table}
\section{THEORETICAL FIDELITY OF DETUNED GATES}\label{sec:sup_optimal_phase}
\begin{figure}
\centering
  \includegraphics[width=1\columnwidth]{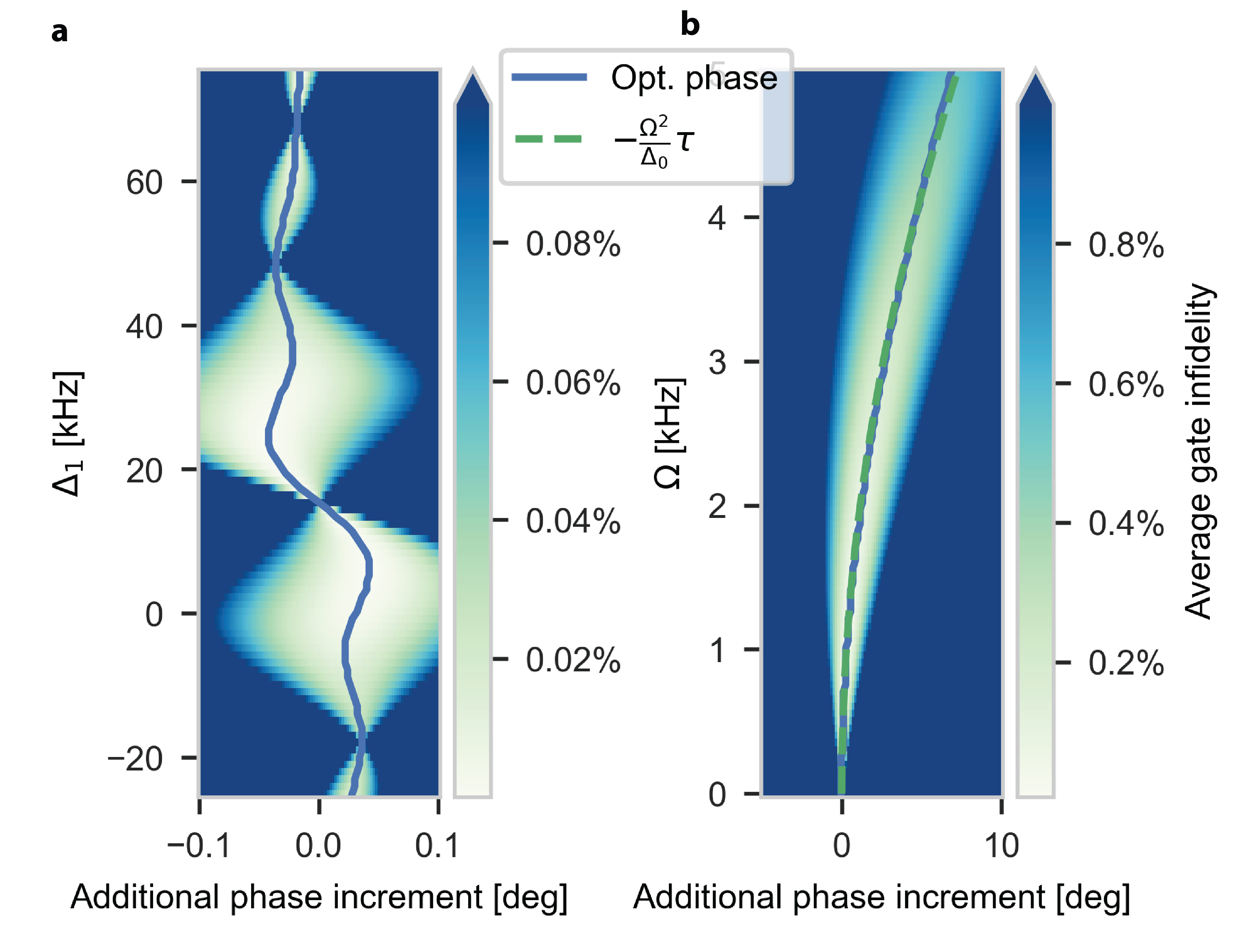}
  \caption{\emph{Optimal single-pulse phase increment}. a) Average gate infidelity of a DDRF gate, over a range of RF detunings $\Delta_1$ and phase increments $\df$. The phase increment on the $x$-axis is added to the resonant $\df$ as derived in Eq. \ref{eq:phase_update_rule}. The optimal phase increment (that yielding the highest fidelity) varies with the applied RF frequency. b) As a function of Rabi frequency $\Omega$, the optimal phase increment for a gate on-resonance with one of the nuclear spin frequencies (here $\Delta_1=0$) can be predicted using the AC-Stark shift. 
  \label{fig:bloch_siegert}
  }
\end{figure}
We present a closer study of the theoretically achievable fidelity of DDRF gates on a single nuclear spin, neglecting $A_\perp$. We consider the spin and gate parameters from Fig. \ref{fig:detuned_ddrf}c: $\Delta = -30.7$kHz, $\tau=24.654$us, $\Omega=313$Hz. We numerically calculate the DDRF gate unitary $U$ using $H_{\mathrm{RF}}$ (Eq. \ref{H_rot}) and compare it to the ideal unitary \begin{align}
    U_{\mathrm{ideal}} = \ket{0}\bra{0} \otimes R_x(\pi/2) + \ket{1}\bra{1} \otimes R_x(-\pi/2) \,.
\end{align} To compare $U$ and $U_{\mathrm{ideal}}$, we decompose $U$ into its rotation angle $\theta$, and rotation axes $\mathbf{\hat{n}}_0$ and $\mathbf{\hat{n}}_1$, for each starting electron spin state (See more detail in \ref{sec:local_window}). We then calculate \begin{align}
    U' = \ket{0}\bra{0} \otimes e^{-i \frac{\pi}{2} \mathbf{\hat{n}}_0\cdot \vec{\sigma}} + \ket{1}\bra{1} \otimes e^{-i \frac{\pi}{2} \mathbf{\hat{n}}_1\cdot \vec{\sigma}}\, ,
\end{align} and compare that to $U_\mathrm{ideal}$ (Fig. \ref{fig:bloch_siegert}). This assumes that the rotation angle of the gate can be tuned to be $\pi/2$, which could experimentally be achieved with a calibration sequence. Tuning the rotation angle with $N$ has no effect on $\mathbf{\hat{n}}_{0,1}$, and fine tuning with $\Omega$ would only slightly affect $\mathbf{\hat{n}}_{0,1}$.

First we consider resonant and off-resonant DDRF gates (Fig. \ref{fig:bloch_siegert},a). We find that high gate fidelities ($F>0.999$) are achievable for a wide range of detunings. To maximise gate fidelity, it is necessary to adjust the single-pulse phase increment $\df$ slightly from the previously derived resonance condition (Eq. \ref{eq:phase_update_rule}). 

Increasing the Rabi frequency reveals why the optimal $\df$ changes (Fig. \ref{fig:bloch_siegert}b). When driving on resonance ($\Delta_1=0$), the optimal phase increment is shifted by \begin{equation} \label{eq:phase_update_BS}
    \delta_{\mathrm{AC}} = - \frac{\Omega^2}{\Delta_0}\tau\, .
\end{equation}
We attribute this effect to the AC-Stark shift\cite{vandersypen_nmr_2005a}. The presence of RF radiation at the $\omega_1$ frequency causes the $\omega_0$ frequency of the nuclear spin to shift, resulting in a different amount of phase being picked up by the spin while the electron is in the $\ket{0}$ state. For other RF frequencies a combined effect of shifting both nuclear spin resonance frequencies occurs.

\section{EXPRESSION FOR THE SENSITIVITY} \label{sec:sup_sensitivity}
We define the single spin sensitivity according to Ref. \cite{degen_quantum_2017}:
\begin{equation} \label{eq:sensitivity_original}
    \sens = \frac{e^{\chi(t)} \sqrt{t + t_m}}{\gamma C(t_m)t} \,,
\end{equation}
where $t$ is the single experiment sensing time, $t_m$ is the readout time, $C$ is a readout efficiency parameter, $\gamma$ is the signal transduction parameter and $\chi(t)$ is the coherence function of the sensor spin. For simplicity, We assume an ideal, instantaneous readout ($C=1, t_m =0$). For the system under study here, this assumption is reasonable as the (single-shot) readout fidelity is $\gtrapprox 90 \%$ and the readout time ($t_m <50 $ \unit{\micro \second}) is significantly shorter than the typical sensing time ($t \sim 1$ \unit{\milli \second}) \cite{bradley_tenqubit_2019}. In our case, $\gamma$ is equal to the (effective) Rabi frequency $\effrabi/(2\pi)$ and has units $\unit{\hertz}/ \mathrm{spin}$, as a single nuclear spin induces phase on the sensor at this rate. Eq. \ref{eq:sensitivity_original} then reduces to:
\begin{equation}
    \sens = \frac{2\pi e^{\chi(N, t)}}{\tilde{\Omega}_{\mathrm{max}}(\Delta,N) \sqrt{t}} \, ,
\end{equation}
which is equal to Eq. \ref{eq:sensitivity} in the main text.

Note that the electron coherence function $\chi(N,t)$ during the DDRF sequence is dependent on the number of applied decoupling pulses $N$ \cite{abobeih_onesecond_2018}:
\begin{equation}
    \chi(N,t) = \left(\frac{t}{T(N)}\right)^{n} \,,
\end{equation}
with $n = 2$, and the coherence time $T(N)$ given by:
\begin{equation}
    T(N) = T_{N=4} \left(\frac{N}{4}\right)^\eta \,,
\end{equation}
with $T_{N=4} = 2.99$ \unit{\milli \second} and $\eta = 0.799$ \cite{abobeih_onesecond_2018}.

\section{OPTIMAL RF DETUNING} \label{sec:sup_optimal_rf_frequency}

The optimal RF detuning condition for a specific $\Delta$ and $\tau$, is found by choosing $\Delta_1$ (and corresponding $\Delta_0 = \Delta_1 - \Delta$) so that Eq. \ref{eq:eff_rabi} is maximised. As Eq. \ref{eq:eff_rabi} and its derivatives are transcendental, it is not trivial to find these maxima. Therefore, we look for an approximate solution, by considering the function's behaviour in the limit of long and short $\tau$.

\begin{figure}[ht]
\centering
  \includegraphics[width=1\columnwidth]{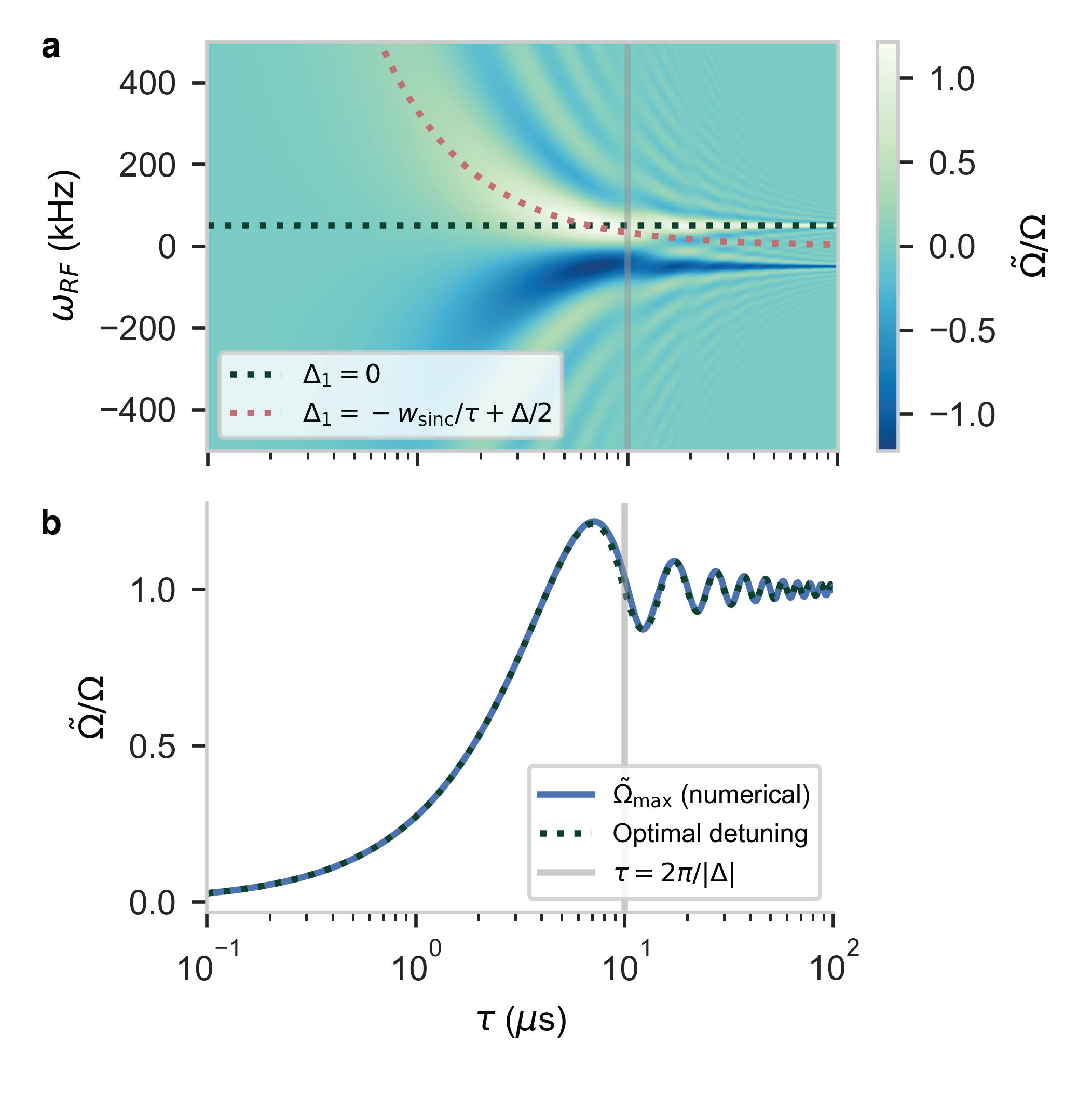}
  \caption{\emph{Optimal RF detuning} a) Evaluation of Eq. \ref{eq:eff_rabi} for a range of interpulse spacings $\tau$, sweeping the RF detuning (by varying $\wrf$). The simulated nuclear spin has a hyperfine coupling $\Delta = \SI{50}{\kilo \hertz}$. The darkgreen dotted line indicates resonant driving, while the red dotted line indicates the (optimal) RF detuning in the regime $\tau < 1/\Delta$ (grey solid line). b) Maximum (relative) effective Rabi frequency as a function of $\tau$, computed by taking the maximum value in (a) (blue solid line). We find good agreement with the analytical (maximum) expression, obtained by evaluating Eq. \ref{eq:eff_rabi} with RF detuning described by Eq. \ref{eq:optimal_detuning} (blue dotted line).
  \label{fig:optimal_detuning}
  }
\end{figure}

In the limit of long $\tau$ (defined as $\tau \gg 2\pi/ \abs{\Delta} $), the optimal detuning is (trivially) $\Delta_1 = 0$. In this limit, the two sinc functions in Eq. \ref{eq:eff_rabi} are completely separated and the global maximum is simply the maximum of the sinc function centered around $\omega_1$ (Fig. \ref{fig:optimal_detuning}a). 

However, for short $\tau$, the spectral width of the sinc functions increases, so that they interfere destructively, which pushes the optimal detuning condition outwards (Fig. \ref{fig:optimal_detuning}a). In the short $\tau$ limit ($\tau \ll 2\pi/\abs{\Delta} $), we can conceptually see that maximum $\effrabi$ will be attained when the difference in driving between the $\omega_0$ and $\omega_1$ transitions is largest, as this maximises the conditional character of the gate. For the square RF pulses considered in this work, this condition is satisfied when the sinc pulse envelope has maximum gradient at $\bar{\omega} = (\omega_1 + \omega_0)/2 $ (schematically illustrated in Fig \ref{fig:sensing_optimisation}a). By evaluating the second derivative of the sinc function, we find that the (maximum gradient) inflexion point is located at a distance $\ws$ away from the peak center, with $\ws\approx 2.082$ the first root of the second derivative of the sinc function. Requiring this point to be positioned precisely in between the two transitions (i.e. $\Delta/2$ away from $\omega_1$), we arrive at the condition:
\begin{equation}
    \Delta_1 = - \frac{\ws}{\tau} + \Delta / 2 \, .
\end{equation}

It is not a priori obvious that these two limiting cases perform well in describing the optimal condition in the intermediate regime ($\tau \sim 2\pi/\abs{\Delta}$). To evaluate the validity of the limiting cases, and to investigate their performance in the intermediate regime, we compute Eq. \ref{eq:eff_rabi} for a range of $\tau$-values (Fig. \ref{fig:optimal_detuning}a) and extract numerically its maximum value (Fig. \ref{fig:optimal_detuning}b, solid blue line). We compare this numerical value to a composite function, generated by joining the limiting descriptions at condition $\tau = 2 \pi/\abs{\Delta}$: 
\begin{equation}\label{eq:sup_optimal_detuning}
  \Delta_1 = \begin{cases}
    - \frac{\ws}{\tau} + \Delta / 2, & \text{if $\tau \lesssim 2\pi/\abs{\Delta}$}.\\
    0, & \text{otherwise}
  \end{cases}
\end{equation}
which is equal to Eq. \ref{eq:optimal_detuning} in the main text. We find good agreement between the numerical maximum, and the effective Rabi frequency obtained by inserting Eq. \ref{eq:sup_optimal_detuning} in Eq. \ref{eq:eff_rabi}, which yields:
\begin{equation} \label{eq:sup_max_rabi_analytical}
    \effrabi / \wR =\begin{cases}
    \mathrm{sinc}\left(\ws - \frac{\Delta \, \tau}{2} \right) - \mathrm{sinc}\left(\ws + \frac{\Delta \, \tau}{2} \right),
    & \text{if $\tau \lesssim 2\pi/\abs{\Delta}$}\\
        
    1 - \mathrm{sinc}(\Delta \, \tau) \, ,
     & \text{otherwise}
  \end{cases}
\end{equation}

We observe only a $< 5\% $ deviation between Eq. \ref{eq:sup_max_rabi_analytical} and the maximum of Eq. \ref{eq:eff_rabi} across the full range (Fig. \ref{fig:optimal_detuning}b, blue dotted line). Thus, we conclude that Eq. \ref{eq:sup_optimal_detuning} is suitable for finding the optimal RF detuning in a sensing setting.

\section{SENSING OPTIMISATION PROCEDURE} \label{sec:sup_sensing_optimisation_procedure}
\begin{figure}
  \includegraphics[width=1\columnwidth]{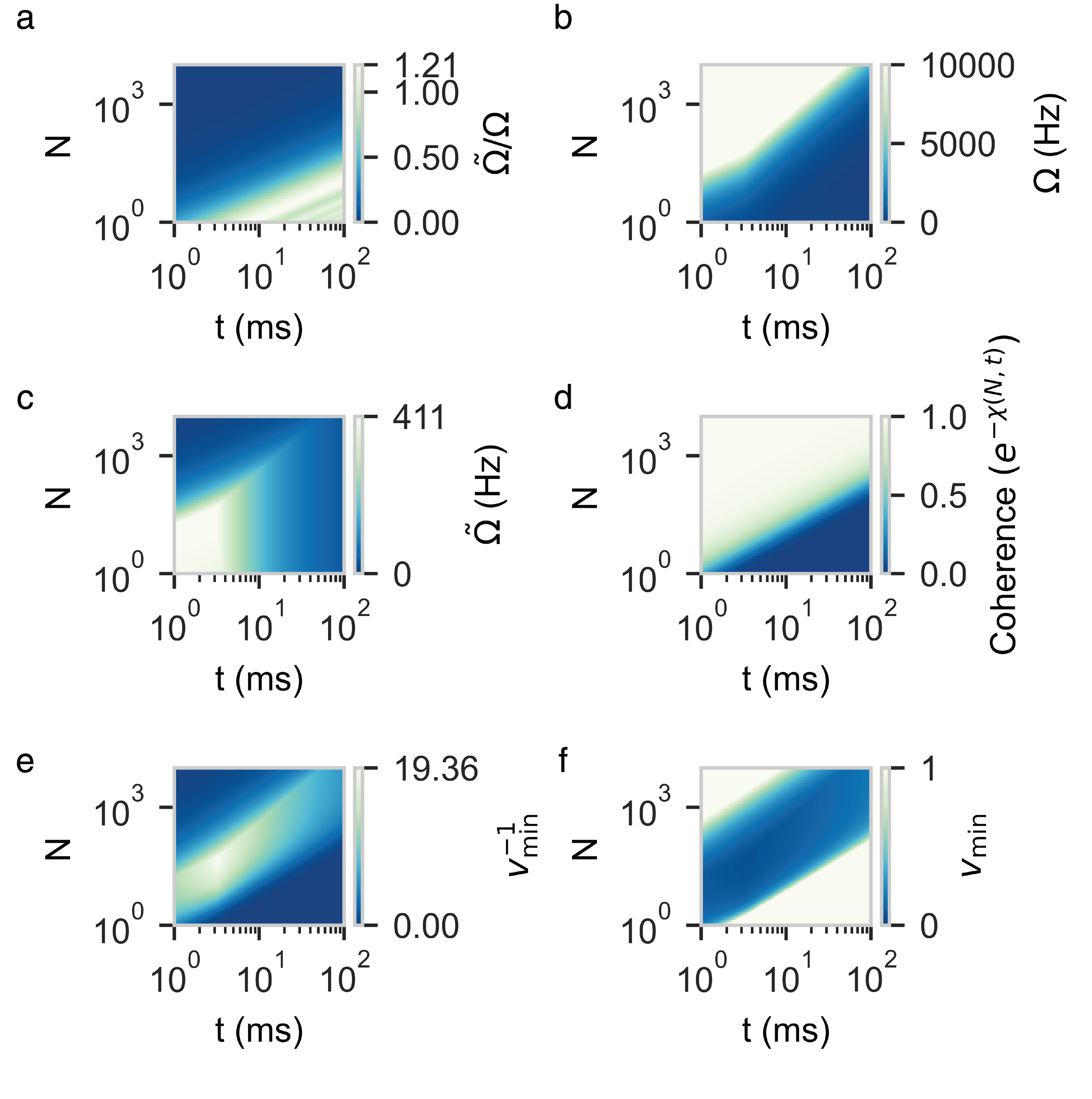}
  \caption{\emph{Optimal parameters for sensing.} a) Relative effective Rabi frequency as a function of the total DDRF sequence length ($t$) and number of applied electron pulses ($N$), obtained by evaluating Eq. \ref{eq:sup_max_rabi_analytical}. b) Physical RF amplitude $\wR$, which is set to counter the suppression of the Rabi frequency visible in (a). c) The effective Rabi frequency, calculated by taking the product of (a) and (b). d) The electron coherence function, taken from Ref. \cite{abobeih_onesecond_2018}. Adding more pulses (shortening the interpulse spacing $\tau$) leads to increased electron coherence at constant gate time $t$. e) The inverse of the sensitivity, defined in Eq. \ref{eq:sensitivity} in the main text, calculated by multiplying (c) and (d). f) The sensitivity function, showing a significant region where single-spin selectivity ($v_\mathrm{min}<1$) is attainable. To generate the data in Fig. \ref{fig:sensing_optimisation}d, we generate this plot for a range of $\Delta$ values and numerically pick the minimum value.
  \label{fig:optimal_N_tau}
  }
\end{figure}

To find optimal sensing parameters we follow the steps described in Fig. \ref{fig:optimal_detuning} and in the main text. First, we compute the effective Rabi frequency for a range of realistic $N$ and $\tau$ values \cite{abobeih_onesecond_2018}. Then, we compute the optimal physical RF amplitude $\wR$. Ideally, this power is increased to exactly counter the reduction in Rabi frequency. However, this requires setting the Rabi amplitude to very large values for small $\tau$. As Eq. \ref{eq:eff_rabi} is only valid in the regime $\Omega \tau \ll 1$, we limit the amplitude to: $\wR = \min\left[1/(2\tau), \SI{10}{\kilo \hertz} \right]$. Figure \ref{fig:optimal_N_tau}b shows the set RF amplitude under this limitation. In the bottom-right region of the graph, the modification of the Rabi frequency is small (see Fig. \ref{fig:optimal_N_tau}a), and therefore $\wR$ is not increased significantly. In the top-left corner however, the desired RF amplitude exceeds our set limits, and its values is capped at \SI{10}{\kilo \hertz}. Note that the optimal sensing parameters depend on the precise choice of the limits. Therefore, it will be important to re-evaluate these in practice, depending on the specific limitations of the experimental setup (e.g. RF delivery efficiency or cryogenics). 

Next, the effective Rabi frequency is computed by multiplying the values obtained in Fig. \ref{fig:optimal_N_tau}a and b. Then, we compute the expected electron coherence for all values of $N$ and $\tau$, based on the decoherence function from Ref. \cite{abobeih_onesecond_2018}. We compute the sensitivity according to Eq. \ref{eq:sensitivity} and plot its value, and its inverse, in Fig \ref{fig:optimal_N_tau}f and e, respectively. A sensitivity below one (blue region in Fig. \ref{fig:optimal_N_tau}f) means the sequence is sensitive to a single nuclear spin.

Finally, we generate these plots for a number of coupling strengths $\Delta$, ranging from \SI{30}{\hertz} to \SI{10}{\kilo \hertz}, and numerically select the maximum value for the sensitivity. Then, we repeat the process without allowing for detuned driving, which results in significantly lower sensitivity. The results are shown in Fig. \ref{fig:sensing_optimisation} in the main text.

\section{BOUND FOR GATE SELECTIVITY} \label{sec:app_gate_sel}
We present a brief mathematical justification for the DDRF gate selectivity that accumulates over the duration of the gate (\ref{eq:phase_sel}).
If we assume negligible driving of the nuclear spin when the electron is in the $\ket{0}$ state (i.e. assuming $\Delta_0 \tau \gg \pi$, which also implies $\Delta_0 \gg \Omega$, given that $\Omega \tau \ll 1$ ), the dynamics can be approximated by the Hamiltonian:
\begin{multline}
    H_{\mathrm{RF, \ket{1}}} = \ket{0}\bra{0} \otimes \Delta_0 I_z + \\\ket{1}\bra{1} \otimes (\Delta_1 I_z + \Omega (\cos \phi I_x + \sin \phi I_y))\,.
\end{multline}
A compact expression for the electron spin's spectroscopy response $\left< S_x\right>$ can be found when the phase increment $\df = -\Delta_0\tau + \pi$ (\ref{sec:local_window}):
\begin{align}\label{eq:spin_response}
    \left<S_x\right> = \frac{1}{2} \left(1 - \frac{2}{1 + \Delta_1^2 / \Omega^2} \sin^2 \left(\frac{N\tau}{2} \sqrt{\Omega^2 + \Delta_1^2}\right)\right).
\end{align}
Note that this $\df$ is the optimal phase increment when $\Delta_1 = 0$ (Eq. \ref{eq:phase_update_rule}). For $\Delta_1 \neq 0$, and for $\ket{0}$ being the electron $m_s=0$ state (giving all spins the same $\Delta_0$), the above equation predicts the electron spin's response to a bystander spin, which diminishes due to the mismatch between $\df$ and the bystander spin's actual evolution. The minimum detuning $\Delta_1$ between the bystander spin and the target spin which causes no crosstalk is given by the first zero of the electron spin's response. Considering the case of an entangling gate between the electron and target spin ($\Omega N \tau = \pi/2$), there is no crosstalk if
\begin{align}
    \Delta_1 = \frac{\sqrt{15}\pi}{2N\tau}.
\end{align}
Translating this to a difference in mean frequency results in 
\begin{align}
    \frac{\Delta_0^t + \Delta_1^t}{2} - \frac{\Delta_0^b + \Delta_1^b}{2} &= \\\wmean^{\mathrm{t}} - \wmean^{\mathrm{b}} &= \frac{\sqrt{15}\pi}{4N\tau}.
\end{align}

Furthermore, the selectivity can be argued from the lorentzian factor $\frac{2}{1+\Delta_1/\Omega^2}$ in Eq. \ref{eq:spin_response}. Under the entangling gate condition $\Omega = \frac{\pi}{2N\tau}$, this lorentzian has a full-width at half-maximum of $\frac{\pi}{N\tau}$.%

\section{MULTI-QUBIT REGISTER OPTIMISATION}\label{sec:register_optimisation}
\begin{figure*}[hb]
  \includegraphics[width=1\textwidth]{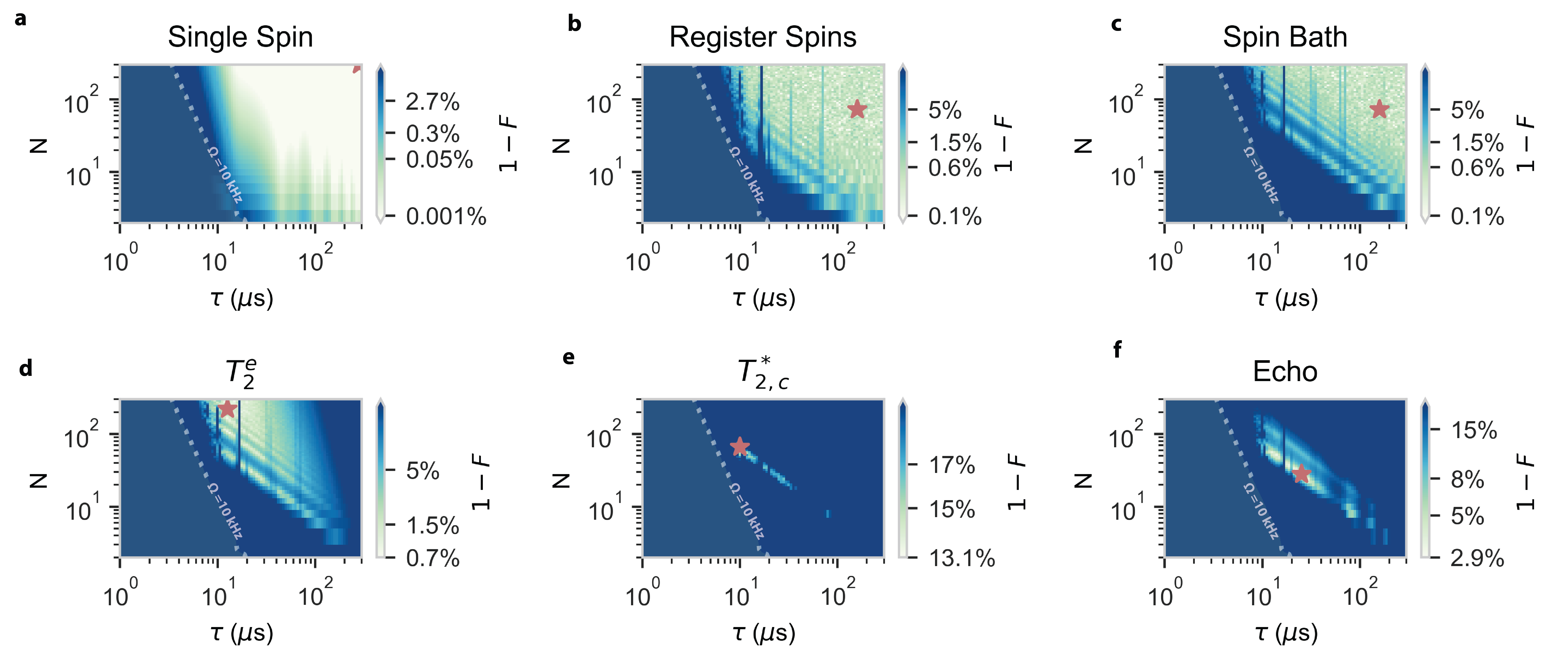}
  \caption{\emph{Infidelity contributions in a 6-qubit 
 register.} $F$ is the 6-qubit gate fidelity. Panel (a-e) show the different contributions in infidelity with $C_4$ as target qubit. a) Considering only the target, for which high-fidelity gates exist across the parameter space. The RF amplitude is limited to $\Omega < \SI{10}{\kilo \hertz}$ (dotted line) resulting in a sharp fidelity drop-off. At larger $N$, the approximation used for obtaining Eq. \ref{eq:eff_rabi} loses validity. b) Considering unitary evolution of all 6 spins in the register (from $\Ucalculated$). c) Also including the signal of a sampled spin bath, using Eq. \ref{eq:nuclear_spin_bath} for $\lambda_{\mathrm{bath}}$ (Eq. \ref{eq:lambda_T2}) and the remaining identified individual bystander spins (Table \ref{tab:a_par}) d) Also including the electron-spin $T_2$ dephasing under dynamical decoupling (Eq. \ref{eq:lambda_T2}), which limits both $\tau$ and gate times. e) Also including the $T_2^\star$ dephasing of the nuclear-spin register qubits by averaging over a distribution of magnetic fields $\delta_B$. This hinders the performance, both by stochastic detuning of the target qubit operation, and by direct dephasing of the register. f) Also including a correction on the phase of all register qubits (Eq. \ref{eq:U_spin_echo}), akin to performing a spin-echo operation on all register qubits, meant to decouple the qubits from quasi-static noise \cite{bradley_tenqubit_2019, abobeih_faulttolerant_2022}.
  \label{fig:sup_alternative_gate_params}
  }
\end{figure*}

In this section we provide more details on the spin register simulations (Fig. \ref{fig:selectivity}). The $M$ two-qubit DDRF unitaries between the electron spin and each nuclear spin in the register are calculated using Eqs. \ref{eq:T0_full} and \ref{eq:T1_full}. The unitaries are corrected for deterministic phase offsets on the idling qubits, which could be taken into account experimentally at no fidelity cost by calibrating virtual-Z gates. The total $(M+1)$-qubit unitary $\Ucalculated$ is subsequently constructed from these two-qubit unitaries, by extending them to the $M+1$ qubit register space and concatenating them, which assumes that they commute. The resulting unitary is an approximation that neglects nuclear-nuclear interactions, as well as electron-nuclear interactions with spins outside the register. Below we will introduce the effects of such interactions in a phenomenological way. We also assume idealised (i.e. instantaneous) electron spin pulses that neglect the effect of the electron-nuclear coupling during the pulse.

The $M+1$-qubit gate fidelity is calculated according to:
\begin{align} 
\label{eq:fidelity_calculation}
    F(\Uideal, \Ucalculated) = \frac{\sum_j \Tr (\Uideal U_j^\dagger \Uideal^{\dagger} \Ucalculated U_j \Ucalculated^{\dagger}) + d^2}{d^2(d+1)} \, ,
\end{align}
where the summation $j$ is over the Pauli matrices and $d$ is the dimension of the Hilbert space \cite{nielsen_simple_2002}.

To incorporate electron-spin dephasing, which commutes with the unitary $\Ucalculated$, a dephasing error channel is applied after $\Ucalculated$. The single-qubit error channel on the electron spin is given by the transformation:
\begin{equation}
    \epsilon(\rho) = \sum_i \hat{K}_i \rho\hat{K}_i^{\dagger} \, ,
\end{equation}
with Kraus operators:
\begin{equation}
\begin{aligned}
    \hat{K}_0 &= \sqrt{\frac{(1+\lambda)}{2}} \, \hat{\mathbb{I}}, \quad & \hat{K}_1 &= \sqrt{\frac{(1-\lambda)}{2}} \, \hat{Z} \, ,
\end{aligned}
\end{equation}
Leading to an error channel for the $M+1$-qubit unitary:
\begin{multline}
        \epsilon_{M+1}(\rho) = \frac{(1+\lambda)}{2}\ide^{\otimes(M+1)}\rho\,{\ide}^{\otimes(M+1)} + \\ \frac{(1-\lambda)}{2}(Z\otimes \ide^{\otimes\,M})\rho (Z\otimes \ide^{\otimes\,M})\ .
\end{multline}
The average gate fidelity of the operator $U_{c}$, followed by the dephasing channel $\epsilon_{M+1}$, is then given by:
\begin{equation}\label{eq:avg_fid}
    F(\Uideal, \Ucalculated, \lambda) = \frac{(1+\lambda)}{2} F(\Uideal, \Ucalculated) + \frac{(1-\lambda)}{2} F(\Uideal, Z'\Ucalculated)\,,
\end{equation}
where $Z' = Z\otimes \ide^{\otimes\,M}$.

The parameter $\lambda$ quantifies the dephasing of the electron spin, which may have one or more independent origins. Considering the dephasing due to the $T_2$ of the electron spin ($\lambda_{T_2}$, due to the nuclear spin bath dynamics) and the direct DDRF-gate-mediated interaction with (mixed) bath spins ($\lambda_{\mathrm{bath}}$), the total dephasing is given by $\lambda = \lambda_{\mathrm{bath}}\lambda_{T_2}$, with

\begin{equation}
\begin{aligned}
\label{eq:lambda_T2}
    \lambda_{\mathrm{bath}} &= \left<\sigma_x\right>_{\mathrm{bath}}, \quad & \lambda_{T_{2}(N, t)} &= e^{-(\frac{t}{T(N)})^n}\, ,
\end{aligned}
\end{equation}
where $\left<\sigma_x\right>_{\mathrm{bath}}$ quantifies the RF-mediated electron dephasing due to the nuclear spin bath (Appendix \ref{sec:app_simulation}) and $T_{2}(N,t)$ is the dephasing time of the electron spin during a dynamical decoupling sequence of $N$ pulses and duration $t$ (Appendix \ref{sec:sup_sensitivity}). 

We implement $T_2^\star$ nuclear-spin dephasing, which does not commute with the unitary evolution during the DDRF gate, by sampling static magnetic fields offsets $\delta_B$ from a Gaussian distribution with standard deviation  $\sigma_B =  1/\left(\sqrt{2}\pi \gamma_cT_2^\star\right)$, where $\gamma_c$ is the $^{13}$C gyromagnetic ratio and $T_2^\star$ the decoherence time for $^{13}$C nuclear spins ($\approx10$ ms). We calculate average gate fidelities (Eq. \ref{eq:avg_fid}) for 10 magnetic fields uniformly sampled within 2 $\sigma_B$. To calculate the final fidelity, we compute an average weighted by the magnetic field probability distribution. 

In typical experimental settings \cite{abobeih_faulttolerant_2022, bradley_tenqubit_2019}, RF spin-echo pulses are performed after the gate to counter nuclear-spin dephasing. We simulate this by explicitly correcting for additional phases acquired by the spins in the register due to the sampled magnetic field offsets. We update the calculated unitary according to: 
\begin{equation}\label{eq:U_spin_echo}
    \Ucalculated' = R_z(- 2 N \tau \gamma_c \, \delta_B )\Ucalculated
\end{equation}

\begin{figure*}[ht]
  \includegraphics[width=1\textwidth]{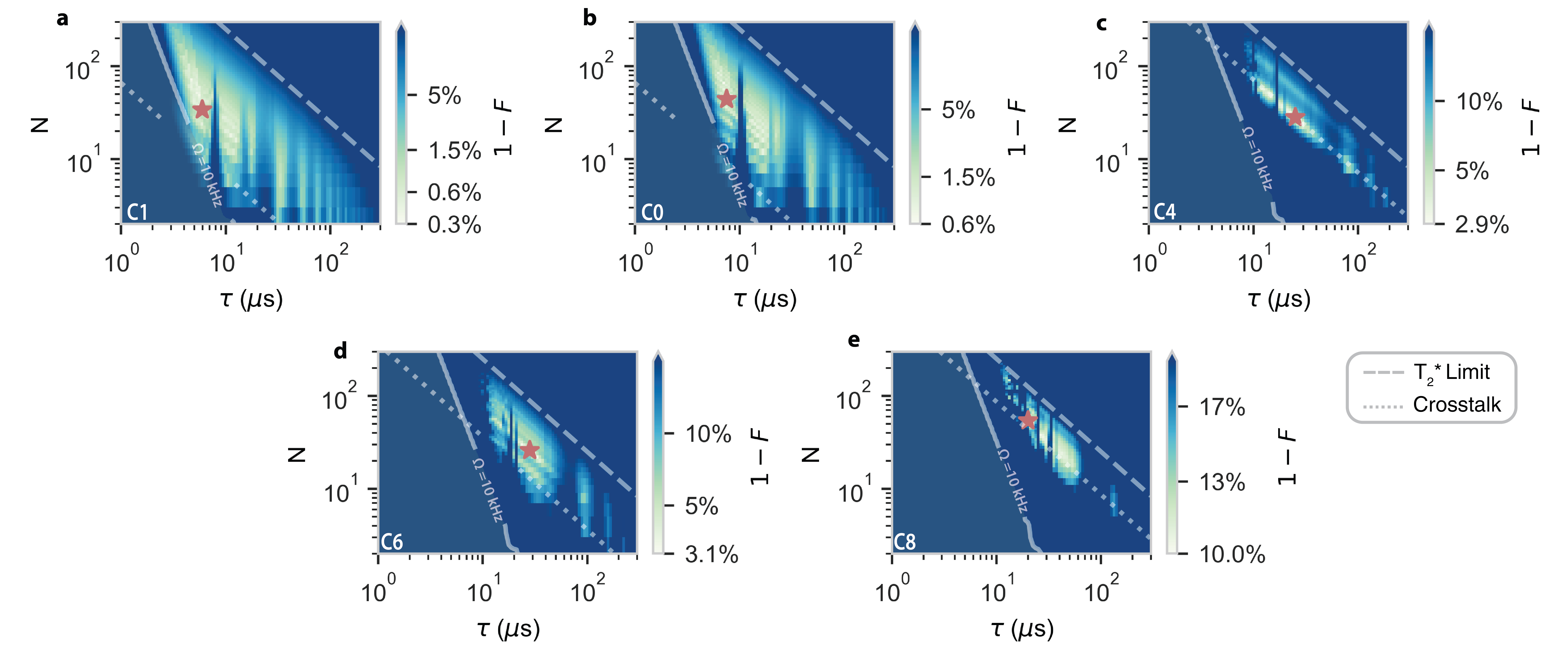}
  \caption{\textit{Characterisation for all 5 electron-nuclear spin controlled gates in the 6-qubit register. 
  \label{fig:sup_five_spins}
  }For each nuclear spin participating in the register we calculate the 6-qubit average gate fidelity for the $\Uideal$ gate, given by Eq. \ref{eq:avg_fid}. The $\tau$ for which crosstalk occurs with other nuclear spins is different for each target spin. The spectral isolation of a spin sets a minimum gate length (dotted lines, Eq. \ref{eq:freq_sel_nv}), and the nuclear spin $T_2^*$ a maximum gate length (dashed lines indicate $2N\tau=5,$ms). For $C_8$ the many spins in close spectral proximity result in a lower attainable gate fidelity.}
\end{figure*}

To give further insight into the effect of each source of infidelity, we investigate their cumulative effects on the infidelity for $C_4$ (Fig. \ref{fig:sup_alternative_gate_params}). We also present the gate fidelity maps for all 5 electron-controlled gates in the 6-qubit register (Fig. \ref{fig:sup_five_spins}).

\clearpage
\newpage
\bibliography{refs}		%

\clearpage
\newpage
\renewcommand{\appendixname}{}

\onecolumngrid
\clearpage
\widetext
\begin{center}
\textbf{\large Supplementary Information for \\``Improved Electron-Nuclear Quantum Gates for Spin Sensing and Control''}
\end{center} 
\renewcommand{\theequation}{S\arabic{section}.\arabic{equation}}
\setcounter{equation}{0}

\renewcommand{\thefigure}{S\arabic{figure}}
\setcounter{figure}{0}

\renewcommand{\thesection}{Supplementary Note \arabic{section}}
\setcounter{section}{0}

\section{Analytic Description of DDRF Gates}\label{sec:local_window}
volution during a DDRF gate, when RF driving is neglected when the electron is in $\ket{0}$. Although this approximation is only valid in the regime where $\Delta > 1/\tau$, it provides much insight in the spin dynamics for the general $\Delta < 1/\tau$ case.

The electron-nuclear Hamiltonian, only considering driving the electron $\ket{1}$ state, in the rotating frame of the RF radiation, is given by:
\begin{align}
    H_{\mathrm{RF, \ket{1}}} = \ket{0}\bra{0} \otimes \Delta_0 I_z + \ket{1}\bra{1} \otimes (\Delta_1 I_z + \Omega (\cos \phi I_x + \sin \phi I_y)),
\end{align}%
The following derivation shows similarities to the derivation used to analyse resonant DD control of nuclear spins \cite{taminiau_detection_2012}. In this approximation of driving during only one electron state, there are two differences between DD and DDRF: the Rabi frequency $\Omega$ takes the role of $A_\perp$ in resonant DD, and the phase of the RF driving can be changed between pulses, while for DD $A_\perp$ always acts along the same axis in the lab frame.

The unitary operator $U_\mathrm{N=2}$ that describes the action of a single (N=2) DDRF block is given by
\begin{align}
    U_\mathrm{DDRF, N=2} = e^{-i H_{\mathrm{RF}} \tau} (R_x(\pi) \otimes R_z(\df))e^{-2 i H_{\mathrm{RF}} \tau} (R_x(\pi) \otimes R_z(\df))e^{-i H_{\mathrm{RF}} \tau} \,,
\end{align}
where $R_z(\df) = e^{-i \df \sigma_z/2}$ are the phase jumps of the RF radiation in between pulses, represented as z-rotations of the nuclear-spin qubit's rotating frame, and $R_x(\pi)$ are rotations of angle $\pi$ around the x-axis of the electron spin, representing the decoupling pulses. 

Due to the block-diagonal nature of the driving Hamiltonian we can define $e^{-i H_{\mathrm{RF}} \tau} = \ket{0}\bra{0} \otimes T_0 + \ket{1}\bra{1} \otimes T_1$, with the unitary operators $T_i$ describing the nuclear spin evolution if the electron spin is in state $\ket{i}$.  A DDRF gate with $2$ decoupling pulses can then be written as: \begin{align} \label{eq:V0V1}
    U_\mathrm{DDRF, N=2} &= -\ket{0}\bra{0}\otimes T_0 R_z(\df) T_1^2 R_z(\df) T_0 - \ket{1}\bra{1}\otimes T_1 R_z(\df) T_0^2 R_z(\df) T_1 \\ &= -\ket{0}\bra{0}\otimes V_0 - \ket{1}\bra{1}\otimes V_1 \,,
\end{align} which is a similar form as obtained for DD in Ref. \cite{taminiau_detection_2012}. Any DDRF gate with more pulses can be found from $(U_\mathrm{DDRF, N=2})^{N/2}$. The operator $V_0$ ($V_1$) is the unitary evolution of the nuclear spin during one DDRF gate $N=2$ unit cell with the electron spin initially in $\ket{0}$ ($\ket{1}$).

Again analogous to Ref. \cite{taminiau_detection_2012}, $V_0$ and $V_1$ can be interpreted as rotations of the nuclear spin under an angle $\theta_0 = \theta_1 = \theta$, and around axes $\mathbf{\hat{n}}_0$ and $\mathbf{\hat{n}}_1$: $V_i = e^{-i \theta/2 \mathbf{\hat{n}}_i\cdot \vec{\sigma}}$, with $\vec{\sigma} = \{\sigma_x, \sigma_y, \sigma_z\}$.
The rotation angle and the inner product of the rotation axes are given by
\begin{align}
    \cos \theta &= \cos (\Omega_{\mathrm{rot}} \tau) \cos (\Delta_0 \tau + \df) - \frac{\Delta_1}{\Omega_{\mathrm{rot}}}\sin (\Omega_{\mathrm{rot}} \tau) \sin (\Delta_0 \tau + \df)\label{eq:cos_theta}\\
    1 - \mathbf{\hat{n}}_0 \cdot \mathbf{\hat{n}}_1 &= \frac{\Omega^2}{\Omega_{\mathrm{rot}}^2} \frac{(1 - \cos (\Omega_{\mathrm{rot}} \tau) )(1 - \cos (\Delta_0 \tau + \df))}{1 + \cos \theta}, \label{eq:rotation_axes}
\end{align}
where $\Omega_{\mathrm{rot}} = \sqrt{\Omega^2 + \Delta_1^2}$. For larger $N$, the solution is given by the same rotation axes, but by a rotation angle $\Theta = N\theta/2$.

The key difference with the analytical result for DD gates is that the resonance condition for an entangling gate can be fulfilled by setting $\df$ appropriately, whereas for DD only the decoupling time $\tau$ is used. Tracking the nuclear spin evolution by setting $\df = -\Delta_0 \tau + \pi$, we find:
\begin{align}
    \cos \theta &= -\cos \Omega_{\mathrm{rot}} \tau\\
    1 - \mathbf{\hat{n}}_0 \cdot \mathbf{\hat{n}}_1 &= \frac{2}{1 + \Delta_1^2 / \Omega^2}.
\end{align}
A fully entangling gate (e.g. $\mathbf{\hat{n}}_0 \cdot \mathbf{\hat{n}}_1 = -1, N \theta/2 = \pi/2$) can be performed when $\Delta_1 = 0, \Omega=\frac{\pi}{2N\tau}$.

Due to the ($S_z I_x$) nature of the interaction, it is useful to consider an experiment where the electron is prepared in $\ket{x} = \frac{1}{\sqrt{2}}(\ket{0}+\ket{{1}})$ and a DDRF sequence is applied. With the nuclear spin in a mixed state, the probability for the electron spin to remain in the $\ket{x}$ state is given by $P_x =\left<S_x\right> + \frac{ 1}{2}$, with \cite{taminiau_detection_2012}:
\begin{align}
    \left<S_x\right> = \frac{1}{2} \left(1 - (1 - \mathbf{\hat{n}}_0 \cdot \mathbf{\hat{n}}_1)\sin^2 \frac{N\theta}{2}\right)
\end{align}
which, under the resonance condition $\df = -\Delta_0 \tau + \pi$, is equal to:
\begin{align}\label{eq:local_window_supp}
    \left<S_x\right> = \frac{1}{2} \left(1 - \frac{2}{1 + \Delta_1^2 / \Omega^2} \sin^2 \left(\frac{N\tau}{2} \sqrt{\Omega^2 + \Delta_1^2}\right)\right).
\end{align}

Alternatively, $\df$ can be set to optimally drive a target spin for $\Delta_1 \neq 0$, tracking its phase evolution during the gate. The solution is $\df \approx -\Delta_1 \tau - \Delta_0 \tau + \pi$ (same as Eq. \ref{eq:phase_update_rule}, see \ref{sec:sup_eff_rabi_derivation} for a derivation). In this case, Eq. \ref{eq:cos_theta} can be approximated in the limit of small rotations per RF pulse ($\Omega \tau \ll 1$) to yield:
\begin{equation}
    \cos{\theta} = - \left( 1 - \frac{1}{2} \left( \,\Omega \, \tau \, \mathrm{sinc}(\Delta_1 \tau) \,\right)^2 \right) + O\left(\Omega^4 \tau^4\right) \, ,
\end{equation}
so that each $N=2$ DDRF block induces a rotation of the nuclear spin by:
\begin{equation}
    \theta \approx \Omega \, \tau \, \mathrm{sinc}(\Delta_1 \tau) + \pi \, ,
\end{equation}
This matches the effective Rabi frequency in the main text(Eq. \ref{eq:eff_rabi}), in the limit considered here where the driving in the $\ket{0}$ state is neglected ($\Delta_0\tau \ll 1$) and demonstrates the RF pulse bandwidth due to $\tau$. The more general case is derived in the next section (\ref{sec:sup_eff_rabi_derivation}).

\clearpage

\section{Effective Rabi Frequency Derivation} \label{sec:sup_eff_rabi_derivation}

As additional justification to the equation for $\tilde{\Omega}$ (Eq. \ref{eq:eff_rabi}), next to the experimental data in the main text, we present a derivation that shows that the DDRF gate can be approximated by a rotation with a reduced Rabi frequency.

The matrix exponentials $T_0$ and $T_1$, as defined in section \ref{sec:local_window}, can be explicitly calculated:
\begin{align}
T_0 = \cos (\tau \sqrt{\Delta_0^2 + \Omega^2}/2) \ide - i \Omega \tau/2 \ \mathrm{sinc} (\tau \sqrt{\Delta_0^2 + \Omega^2}/2) \sigma_x - i \Delta_0 \tau/2 \ \mathrm{sinc} (\tau \sqrt{\Delta_0^2 + \Omega^2}/2) \sigma_z,\label{eq:T0_full}\\
T_1 = \cos (\tau \sqrt{\Delta_1^2 + \Omega^2}/2) \ide - i \Omega \tau/2 \ \mathrm{sinc} (\tau \sqrt{\Delta_1^2 + \Omega^2}/2) \sigma_x - i \Delta_1 \tau/2 \ \mathrm{sinc} (\tau \sqrt{\Delta_1^2 + \Omega^2}/2) \sigma_z,\label{eq:T1_full}
\end{align}
where $\sigma_x$ and $\sigma_z$ are the Pauli spin matrices. This form allows an efficient and accurate numerical calculation of the unitary operator of the DDRF gate (as only multiplication of $2\times2$ matrices is required), which was used to calculate the individual spin signals in Fig. \ref{fig:detuned_ddrf}b, Fig \ref{fig:selectivity}d (Appendix \ref{sec:app_simulation}). 

Directly calculating $V_0$ or $V_1$ results in an analytically complex expression that is not easily simplified. Progress can be made by considering the limit $\Omega\tau \ll 1$. This is justified as DDRF is designed to target weakly-coupled nuclear spins ($\Delta < 1/T_{2,e}^*$) for which decoupling of the electron spin is needed, breaking up the rotation of the nuclear spin into small amounts.

To first order in $\Omega \tau$, $U_\mathrm{DDRF, N=2}$ (Eq. \ref{eq:V0V1}) can be approximated by\begin{align*}
V_0 &= \ide \left[\cos (\tau  (\text{$\Delta _0$}+\text{$\Delta_1 $})+\df ) + O\left(\Omega^2 \tau^2\right)\right]\\
&- \sigma_x\left[\frac{i \Omega \tau  (\text{$\Delta_1 $} (\sin (\tau  (\text{$\Delta_0 $}+\text{$\Delta_1
   $})+\df )-\sin (\text{$\Delta_1 $} \tau +\df ))+\text{$\Delta_0 $} \sin (\text{$\Delta_1
   $} \tau ))}{\text{$\Delta_0 $} \text{$\Delta_1 $} \tau } + O\left(\Omega^3 \tau^3\right)\right]\\
&- \sigma_z \left[i \sin (\tau (\Delta_0 + \Delta_1) + \df)+ O\left(\Omega^2 \tau^2\right)\right]\\
V_1 &= \ide \left[\cos (\tau  (\text{$\Delta _0$}+\text{$\Delta_1 $})+\df ) + O\left(\Omega^2 \tau^2\right)\right]\\
&- \sigma_x\left[\frac{i \Omega \tau  (\text{$\Delta_0 $} (\sin (\tau  (\text{$\Delta_0 $}+\text{$\Delta_1
   $})+\df )-\sin (\text{$\Delta_0 $} \tau +\df ))+\text{$\Delta_1 $} \sin (\text{$\Delta_0
   $} \tau ))}{\text{$\Delta_0 $} \text{$\Delta_1 $} \tau } + O\left(\Omega^3 \tau^3\right)\right]\\
   &- \sigma_z \left[i \sin (\tau (\Delta_0 + \Delta_1) + \df)+ O\left(\Omega^2 \tau^2\right)\right]\\
\end{align*}

Thus the action of the DDRF gate on the nuclear spin can be approximated by a z-rotation, and an x-rotation by an angle that scales linearly with $\Omega$, as would be expected. The derivation also shows the $\df$ resonance conditions. Setting $\df = -\Delta_0\tau -\Delta_1\tau + \pi$ results in: \begin{align}
    V_{0, \mathrm{con}} &= \ide - \sigma_x i \Omega \tau (\mathrm{sinc} (\Delta_1 \tau) - \mathrm{sinc} (\Delta_0 \tau)) + O\left(\Omega^2 \tau^2\right)\\
    V_{1, \mathrm{con}} &= \ide + \sigma_x i \Omega \tau (\mathrm{sinc} (\Delta_1 \tau) - \mathrm{sinc} (\Delta_0 \tau)) + O\left(\Omega^2 \tau^2\right),
\end{align}
which describes (to first order) an x-rotation conditional on the electron state, with a modified Rabi frequency $\tilde{\Omega}$ described by equation \ref{eq:eff_rabi}.

Setting $\df = -\Delta_0\tau -\Delta_1\tau$ results in the operators \begin{align}
    V_{0, \mathrm{uncon}} = V_{1, \mathrm{uncon}} = \ide - \sigma_x i \Omega \tau (\mathrm{sinc} (\Delta_0 \tau) + \mathrm{sinc} (\Delta_1 \tau)) + O\left(\Omega^2 \tau^2\right),
\end{align}
which describe an unconditional x-rotation. 

One higher order effect neglected in this analysis is the AC-Stark shift \cite{vandersypen_nmr_2005a}, quadratic in $\Omega^2$, which can shift the resonant phase increment $\df$ (see also Appendix \ref{sec:sup_optimal_phase}). Furthermore, in the regime where $\wR \gg \Delta$, the nuclear spin is driven by the RF field regardless of the electron spin state and the conditionality of the interaction is no longer dependent on the set phase increment (Eq. \ref{eq:phase_update_rule}). Instead, the nuclear eigenstates become dressed along the $x$-axis and decoupling the electron results in a perturbation along the $z$-axis (whose magnitude scales with $A_\parallel$). This regime can also be used for spin-selective sensing and control, and forms the basis for the recently developed AERIS protocol \cite{munuera-javaloy_highresolution_2023}

\clearpage

\end{document}